%% file: main.tex
\documentclass[10pt]{article}
\usepackage{amssymb,amsmath}
\usepackage{scrextend}
\usepackage{setspace}
\usepackage{lipsum}
\usepackage{titlesec}
\usepackage{filecontents}

\usepackage[utf8]{inputenc}
\usepackage{booktabs, float, color, colortbl, graphicx, ragged2e, threeparttablex, threeparttable, longtable, placeins, booktabs, float, tabularx}
\usepackage{placeins}
\usepackage[english]{babel}
\usepackage{adjustbox}
\usepackage{acronym}
\usepackage{csquotes}
\usepackage{hyperref}
\usepackage{subcaption}
\usepackage{dsfont}

\usepackage{tikz}
\usepackage{relsize}
\usetikzlibrary{arrows,shapes,positioning,calc}

\hypersetup{plainpages=false,naturalnames=true,bookmarksnumbered=true,bookmarksopen=false,colorlinks=true,urlcolor=black,
	linkcolor=black,filecolor=black,citecolor=black,pdfpagemode=UseOutlines}
\usepackage[natbib,style=authoryear,giveninits=true,maxbibnames=9,maxcitenames=1,uniquelist=false]{biblatex}

\AtEveryBibitem{\clearlist{language}
  \clearfield{language}
  \clearlist{isbn}\clearfield{isbn}
  \clearlist{issn}\clearfield{issn}
  \clearfield{month}
  \iffieldundef{url}{}
      {\iffieldundef{doi}{}
        {\clearlist{url}\clearlist{eprint}}
      }
  \iffieldundef{eprint}{}
      {\iffieldundef{doi}{}
        {\clearlist{eprint}}
      }
} 

\AtBeginBibliography{}

\acrodef{CDER}{Canadian Center for Data Development and Economic Research}
\acrodef{PEI}{Prince Edward Island}
\acrodef{LEAP}{Longitudinal Employment Analysis Program}
\acrodef{BHP}{Establishment History Panel}
\acrodef{BR}{Business Register}
\acrodef{ILU}{Individual Labour Unit}
\acrodef{ALU}{Average Labour Unit}
\acrodef{LBD}{Longitudinal Business Database}
\acrodef{SynLBD}{Synthetic \ac{LBD}}
\acrodef{CBP}{County Business Patterns}
\acrodef{SBUSB}{Statistics of U.S. Businesses}
\acrodef{SEPH}{Survey of Employment, Payrolls and Hours}
\acrodef{LEHD}{Longitudinal Employer-Household Dynamics}
\acrodef{QWI}{{Q}uarterly {W}orkforce {I}ndicators}
\acrodef{BDS}{Business Dynamics Statistics}
\acrodef{NAICS}{North American Industrial Classification System}
\acrodef{SIC}{Standard Industrial Classification}

\newcommand{\SynLBD}{\textsc{SynLBD}}
\newcommand{\sym}[1]{\rlap{#1}}

\usepackage{mathptmx} 
\usepackage{fancyhdr}

\def \mytitle{Applying Data Synthesis for Longitudinal Business Data across Three Countries }

\begin{document}
\thispagestyle{empty}

\begin{onehalfspace}
	\begin{center} 
		{\Large \bf {\mytitle} }
	\end{center}
\end{onehalfspace}

\vspace*{-4mm}

\begin{center}
	\begin{large} 
	{\bf M. Jahangir Alam}\footnote{Department of Applied Economics, HEC Montr\'eal, and Department of Economics, Truman State University. USA. E-mail: jalam@truman.edu.\\ ORCID: https://orcid.org/0000-0001-6478-114X.},
	{\bf Benoit Dostie}\footnote{Department of Applied Economics, HEC Montr\'eal. USA. E-mail: benoit.dostie@hec.ca. ORCID: https://orcid.org/0000-0002-4133-2365.}	 
    {\bf J\"org Drechsler}\footnote{Institute for Employment Research. USA. E-mail: joerg.drechsler@iab.de}	 
    {\bf Lars Vilhuber}\footnote{Cornell University. E-mail: lars.vilhuber@cornell.edu. \\ORCID: https://orcid.org/0000-0001-5733-8932.}	 
    \end{large}
\end{center}

\begin{center}\begin{large} \textbf{ABSTRACT} \end{large} \end{center}

\begin{addmargin}[6mm]{6mm}
\begin{small}
\begin{singlespace}

Data on businesses collected by statistical agencies are challenging to protect. Many businesses have unique characteristics, and distributions of employment, sales, and profits are highly skewed. Attackers wishing to conduct identification attacks often have access to much more information than for any individual. As a consequence, most disclosure avoidance mechanisms  fail to strike an acceptable balance between usefulness and confidentiality protection. Detailed aggregate statistics by geography or detailed industry classes are rare, public-use microdata on businesses are virtually inexistant, and access to confidential microdata can be burdensome. 
Synthetic microdata have been proposed as a secure mechanism to publish microdata, as part of a broader discussion of how  to provide broader access to such data sets to researchers.
In this article, we document an experiment to create analytically valid synthetic data, using the exact same model and methods previously employed for the United States, for data from two different countries: Canada (\ac{LEAP}) and Germany (\ac{BHP}). We assess utility and protection, and provide an assessment of the feasibility of extending such an approach in a cost-effective way to other data.


\end{singlespace}
\end{small}
\end{addmargin}

\smallskip
\acresetall
\section{Introduction}
There is growing demand for firm-level data allowing detailed studies of firm dynamics. Recent examples include \textcite{NBERc0480}, who use cross-country firm-level data to study average post-entry behavior of young firms. \textcite{10.1257/aer.20141280} use the Business Dynamics Statistics (BDS) to show the role of firm size in firm dynamics. However, such studies are made difficult due to the limited or restricted access to firm-level data.

Data on businesses collected by statistical agencies are challenging to protect. Many businesses have unique characteristics, and distributions of employment, sales and profits are highly skewed. Attackers wishing to conduct identification attacks often have access to much more information than for any individual. It is easy to find examples of firms and establishments that are so dominant in their industry or location that they would be immediately identified if data that included their survey responses or administratively collected data were publicly released.  Finally, there are also greater financial incentives to identifying the particulars of some firms and their competitors.

As a consequence, most disclosure avoidance mechanisms  fail to strike an acceptable balance between usefulness and confidentiality protection. Detailed aggregate statistics by geography or detailed industry classes  are rare, public-use microdata on business are virtually inexistant,\footnote{See \citet{NBERw22095} and \citet{startupcartography} for an example of scraped, public-use microdata.} and access to confidential microdata can be burdensome. It is not uncommon that access to establishment microdata, if granted at all, is provided through data enclaves (Research Data Centers), at headquarters of statistical agencies, or some other limited means, under strict security conditions. These restrictions on data access reduce the growth of knowledge by increasing the cost to researchers of accessing the data.

Synthetic microdata have been proposed as a secure mechanism to publish microdata \citep{drechsler2008,RePEc:taf:japsta:v:39:y:2012:i:2:p:243-265,NAP11844,SJIAOS-2014c}, based on suggestions and methods first proposed by \citet{rubin93} and \citet{little93}. Such data are  part of a broader discussion of how  to provide improved access to such data sets to researchers  \citep{Bender2009,Vilhuber2013,AbowdLane2004,AbowdSchmutte_BPEA2015}.\footnote{For a recent overview of some, see \citet{VilhuberAbowdReiter:Synthetic:SJIAOS:2016}. See \citet{dre:2011} for a review of the theory and applications of the synthetic data methodology.
	Other access methods include secure data enclaves (e.g., research data centers of the U.S. Federal Statistical System, of the  German Federal Employment Agency, others), and  remote submission systems. We will comment on the latter in the conclusion.}
For business data, synthetic business microdata were released in the United States \citep{KinneyEtAl2011} and in Germany  \citep{RePEc:iab:iabfme:201101_de} in 2011. The former data set, called \ac{SynLBD}, was  released to an easily web-accessible computing environment \citep{AbowdVilhuber2010}, and combined with a validation mechanism.  By making disclosable synthetic microdata available through a remotely accessible data server, combined with a validation server, the SynLBD approach alleviates some of the access restrictions associated with economic data. The approach is mutually beneficial to both agency and researchers. Researchers can access public use servers at little or no cost, and can later validate their model-based inferences on the full confidential microdata. Details about the modeling strategies used for the SynLBD   can be found in  \citet{KinneyEtAl2011} 
and \citet{RePEc:cen:tnotes:11-01}.

In this article, we document an experiment to create analytically valid synthetic data, using the exact same model and methods previously used to create the \ac{SynLBD}, but applied to data from two different countries: Canada (\ac{LEAP}) and Germany (\ac{BHP}). We describe all three countries' data in Section~\ref{sec:data}.

In Canada, the Canadian Center for Data Development and Economic Research (CDER) was created in 2011 to allow Statistics Canada to make better use of its business data holdings, without compromising security. Secure access  to business microdata for approved analytical research projects is done through a physical facility located in Statistics Canada’s headquarters. 

CDER implements many risk mitigation measures to alleviate the security risks specific to micro-level business data including limits on tabular outputs, centralized vetting, monitoring of program logs. Access to the data is done through a Statistics Canada designed interface, in which actual observations cannot be viewed. But the cost of traveling to Ottawa remains the most significant barrier to access.

The Institute for Employment Research (IAB) in Germany also strictly regulates the access to its business data. All business data can  be accessed exclusively onsite at the research data center (RDC) and only after the research proposal has been approved by the Federal Ministry of Labour and Social Affairs. All output is carefully checked by staff at the RDC and only cleared output can be removed  from the RDC. 

The experiment described in this paper aims not so much at finding the \textit{best} synthetic data method for each file, but rather to assess the effectiveness of using a `pre-packaged' method to cost-effectively generate synthetic data. In particular, while we could have used newer implementations of methods combined with a pre-defined or automated model \citep{JSSv074i11,Raab_Nowok_Dibben_2018}, we chose to use the exact SAS code used to create the original \ac{SynLBD}. A brief synopsis of the  method, and any adjustments we made to take into account structural data differences, are described in Section~\ref{sec:methodology}.

We verify the analytical validity of the synthetic data files so created along a variety of measures. First, we show how well average firm characteristics (gross employment, total payroll) in the synthetic data  match those from the original data. We also consider how well the synthetic data  replicates various measures of firm dynamics (entry and exit rates) and job flows (job creation and destruction rate). Second, we assess whether measures of economic growth vary between both data sets using dynamic panel data models. Finally, to assess the analytical validity from a more general perspective, we  compute global validity measures based on the ideas of propensity score matching as proposed by \citet{Woo_Reiter_Oganian_Karr_2009,Snoke_RSSA2018}.

To assess how protective the newly created synthetic database is, we estimate the probability that the synthetic first year equals the true first year given the synthetic fist year.

The rest of the paper is organized as follows. Section 2 describes the different data sources and summarizes which steps were taken to harmonize the data sets prior to the actual synthesis. Section 3 provides some background on the synthesis methods, limitations in the applications, and a discussion of some of the measures, which are used in Section 4 to evaluate the analytical validity of the generated data sets. Preliminary results regarding the achieved level of protection are included in Section 5. The paper concludes with a discussion of the implications of the study for future data synthesis projects.
 \newpage
\section{Data} 
\label{sec:data}

In this section, we briefly describe the structure of the three data sources.

\subsection{United States: \acf{LBD}}

The \ac{LBD} \citep{LBD} is created from the U.S. Census Bureau's \ac{BR} by creating longitudinal links of establishments using name and address matching. The database has information on birth, death, location, industry,  firm  affiliation of employer establishments, and ownership by multi-establishment firms, as well as their employment over time, for nearly all sectors of the economy from 1976 through 2015 (as of this writing). It serves as a key linkage file as well as a research data set in its own right for numerous research articles, as well as a tabulation input to the U.S. Census Bureau's \acl{BDS} \citep[\acs{BDS}]{BDS}. Other statistics created from the underlying Business Register include the \acl{CBP} \citep[\acs{CBP}]{CBP} and the \acl{SBUSB} \citep[\acs{SBUSB}]{SBUSB}. For a full description, readers should consult  \citet{RePEc:cen:wpaper:02-17,}. The key variables of interest for this experiment are birth and death dates, payroll, employment, and the industry coding of the establishment.  \citet{SJIAOS-2014d} explore a possible expansion of the synthesis methods described later to include location and firm affiliation. Note that information on payroll and employment does not come from individual-level wage records, as is the case for both the Canadian and German data sets described below, as well as for the \acl{QWI} \citep{AbowdEtAl2009} derived from the \acl{LEHD} \citep[LEHD]{RePEc:cen:wpaper:18-27} in the United States. Thus, methods that connect establishments based on labor flows \citep{BenedettoEtAl2007,RePEc:iab:iabfme:201006_en} are not employed. We also note that payroll is the cumulative sum of wages paid over the entire calendar year, whereas employment is measured as of March 12 of each year.

\subsection{Canada: \acf{LEAP}}

The \ac{LEAP} \citep{StatisticsCanada2019} contains information on annual employment for each employer business in all sectors of the Canadian economy. It covers incorporated and unincorporated businesses that issue at least one annual statement of remuneration paid (T4 slips) in any given calendar year. It excludes self-employed individuals or partnerships with non-salaried participants.

To construct the \ac{LEAP}, Statistics Canada uses three sources of information: (1) T4 administrative data  from the Canada Revenue Agency (CRA), (2) data from Statistics Canada's \acl{BR} \citep{StatisticsCanada2019a}, and (3) data from  Statistics Canada's \acf{SEPH} \citep{StatisticsCanada2019b}. 
In general, all employers in Canada provide employees with a T4 slip if they paid employment income, taxable allowances and benefits, or any other remuneration in any calendar year. The T4 information is reported to the tax agency, which in turn provides this information to Statistics Canada. 
The Business Register is Statistics Canada's central repository of baseline information on businesses and institutions operating in Canada. It is used as the survey frame for all business related data sets.
The objective of the \ac{SEPH} is to provide monthly information on the level of earnings, the number of jobs, and hours worked by detailed industry at the national and provincial levels. To do so, it combines a census of approximately one million payroll deductions provided by the CRA, and the Business Payrolls Survey, a sample of 15,000 establishments.

The core \ac{LEAP}  contains four variables (1) a longitudinal Business Register Identifier (LBRID), (2) an industry classification, (3) payroll and (4) a measure of employment. 
The LBRID uniquely identifies each enterprise and is derived from the Business Register. To avoid ``false'' deaths and births due to mergers, restructuring or changes in reporting practices, Statistics Canada uses employment flows. Similar to \citet{BenedettoEtAl2007} and \citet{RePEc:iab:iabfme:201006_en}, the method  compares the cluster of workers in each newly identified enterprise with all the clusters of workers in firms from the previous year. This comparison yields a new identifier (LBRID) derived from those of the \ac{BR}.
The industry classification comes from the \ac{BR} for single-industry firms. If a firm operates in multiple industries, information on payroll from the \ac{SEPH} is used to identify the industry in which the firm pays the highest payroll. Prior to 1991, information on industry was based on the SIC,  but it is currently based on the  North American Industrial Classification System (NAICS). We use the information at the NAICS four-digit (industry group) level. 
The firm's payroll is measured as the sum of all T4s  reported to the CRA for the calendar year.
Employment is measured either using \ac{ILU} or \ac{ALU}. \acp{ALU} are obtained by dividing the payroll by the average annual earnings in its industry/province/class category computed using the \ac{SEPH}. \acp{ILU} are a head count of the number of T4 issued by the enterprise, with employees working for multiple employers split proportionately across firms according to their total annual payroll earned in each firm.

For the purpose of this experiment,  we exclude the public sector (NAICS 61, 62, and 91), even though it is contained in the database, because it may not be accurately captured \citep{StatisticsCanada2019}. Statistics Canada does not publish any statistics for those sectors. 
\subsection{Germany: \acf{BHP}}

The core database for the Establishment History Panel is the German Social Security Data  (GSSD), which is based on the integrated notification procedure for the health, pension and unemployment insurances,   introduced in  1973. Employers report information on all their employees. Aggregating this information via an establishment identifier yields the Establishment History Panel \citep[German abbreviation: BHP]{BHP}. We used data from  1975 until 2008, which at the time this project started was the most current data available for research. Information for the former Eastern German States is limited to the years 1992-2008. 

Due to the purpose and structure of the GSSD, some variables present in the \ac{LBD} are not available on the  \ac{BHP}. Firm-level information is not captured, and it is thus not known whether establishments are part of a multi-establishment employer. In 1999, reporting requirements were extended to all establishments; prior to that date, only establishments that  had at least one employee covered by social security on the reference date June 30 of each year were subject to filing requirements. Payroll and employment are both based on a reference date of June 30, and are thus consistent point-in-time measures. 
Industries are identified according to the WZ 2003 classification system \citep{WZ2003} at the five digit level.\footnote{The WZ 2003 classification system is compliant with the requirements of the Statistical Classification of Economic Activities in the European Community (NACE Rev. 1.1), which is based on the International Standard Industrial Classification (ISIC Rev. 3.1).} We aggregated the industry information for this project  using the first four digits of the coding system.

\subsection{Harmonizing and Preprocessing}

In all countries, the underlying data provide annual measures. However, \SynLBD{} assumes a longitudinal (wide) structure of the data set, with invariant industry (and location). In all cases, the modal industry is chosen to represent the entity's industrial activity. 
Further adjustments made to the \ac{BHP} for this project include estimating full-year payroll, creating time-consistent geographic information, and applying employment flow methods \citep{RePEc:iab:iabfme:201006_en} to adjust for spurious births and deaths in establishment identifiers. \citet{SJIAOS-2014b} provide a detailed description of the steps taken to harmonize the input data.

In both Canada and Germany, we encountered various technical and data-driven limitations. In all countries, data in the first year and last year are occasionally problematic, and such data  were dropped. 
Both the German and the Canadian data experience some level of industry coding change, which may affect the classification of some entities. Furthermore, due to the nature of the underlying data, entities are establishments in Germany and the US, but employers in Canada. 

After the various standardizations and choices made above, the data structure is intended to be comparable, as summarized in Table~\ref{tab:common_Variable}. The column "Nature" identifies the treatment of the variable in the synthesis process \SynLBD. 

\begin{table}[H]
  \centering\footnotesize
  \caption{Variable descriptions and comparison}  \label{tab:common_Variable}
  
\resizebox{\columnwidth}{!}{\medskip
  \renewcommand{\arraystretch}{1}
 \setlength{\tabcolsep}{4.5pt}
 \begin{tabular}{l  l l c c c c c}
    \toprule
    \textbf{Name}&\textbf{Type} &\textbf{Description} &\textbf{US} & \textbf{Canada} &\textbf{Germany} &\textbf{Nature}\\
    \midrule
Entity Identifier& identifier& & Establishment & Employer & Establishment &Created\\
\midrule
Industry code&Categorical& Various across countries &SIC3 & NAICS4 & WZ2003 &Unmodified\\
             &           &                          &(3-digit )& (4-digit) &(4-digit)  &\\
\midrule
First year&Categorical&First year entity is observed &\multicolumn{3}{c}{--- firstyear ---}&Synthesized\\
Last year&Categorical&Last year entity is observed &\multicolumn{3}{c}{--- lastyear ---}&Synthesized\\
Year&Categorical&Year dating of annual variables&\multicolumn{3}{c}{--- year ---}&Derived\\
\midrule
Employment & Continuous & Employment measure & Count & ALU* & Count & Synthesized \\
            &            &                    & (March 15) &(annual)& (June 30)&\\
Payroll&Continuous&  Payroll (annual)& Reported & Computed & Computed,  &Synthesized\\
       &          &                  &          &          & Adjusted\\
   \bottomrule
  \multicolumn{7}{l}{* ALU = Average Labour Unit. See text for additional explanations.}\\
  \end{tabular} 
}

\end{table}

\section{Methodology}
\label{sec:methodology}

To create a partially synthetic database with analytic validity from longitudinal establishment data, \citet{RePEc:cen:tnotes:11-01} synthesize the life-span of establishments, as well as the evolution of their employment, conditional on industry over that synthetic lifespan. Geography is not synthesized, but is suppressed from the released file \citep{RePEc:cen:tnotes:11-01}. Applying this to the \ac{LBD}, \citet{KinneyEtAl2011}  created the current version of the Synthetic LBD,  based on the Standard Industrial Classification (SIC) and extending through 2000. \citet{RePEc:cen:wpaper:14-12} describe efforts to create a new version of the Synthetic LBD, using a longer time  series (through 2010) and newer industry coding (NAICS), while also adjusting and extending the models for  improved  analytic validity and  the imputation of additional variables. In this paper, we refer to and re-use the older methodology, which we will call \SynLBD. Our emphasis is on the comparability of results obtained for a given methodology across the various applications.

The general approach to data synthesis is to generate a joint posterior predictive distribution of $Y|X$ where $Y$ are variables to be synthesized and $X$ are unsynthesized variables. The synthetic data are generated by sampling new values from this distribution. In \SynLBD, variables are synthesized in a sequential fashion, with categorical variables being generally processed first using a variant of Dirichlet-Multinomial models. Continuous variables are then synthesized using a normal linear regression model with kernel density-based transformation \citep{WOODCOCK20094228}.\footnote{\textcite{RePEc:cen:wpaper:14-12} shift  to a Classification and Regression Trees (CART) model with Bayesian bootstrap. } The synthesis models are run independently for each industry. \SynLBD{} is implemented in SAS\texttrademark, which is frequently used in national statistical offices.

To evaluate whether synthetic data algorithms developed in the U.S. can be adapted to generate similar synthetic data for other countries, \textcite{RePEc:cen:wpaper:14-13} implement \SynLBD{} to the German Longitudinal Business Database (GLBD). In this paper, we extend the analysis from the earlier paper, and extend the application to the Canadian context (SynLEAP). 

\subsection{Limitations}

In all countries, the synthesis of certain industries failed to complete. In both Canada and the US, this number is less than 10. In Canada, they account for about 7 percent of the total number of observations (see Table 13
in the  Appendix).

In the German case, our experiments were limited to only a handful of industries, due to a combination of time and software availability factors. The results should still be considered preliminary. In both countries, as outlined in Section~\ref{sec:data}, there are subtle but potentially important differences in the various variable definitions. Industry coding differs across all three countries, and the level of detail in each of the industry codings may affect the success and precision of the synthesis.\footnote{\textcite{StatisticsCanada1991}, when comparing the 1987 US \ac{SIC} to the 1980 Canadian \ac{SIC},  already pointed out that the degree of specialization, the organization of production, and the size of the respective markets differed. Thus, the density of establishments within each of the chosen categories is likely to affect the quality of the synthesis.} 

As noted in Section~\ref{sec:data}, entities are establishments in Germany and the US, but employers in Canada. \SynLBD{} should work on any level of entity aggregation (see \citet{RePEc:cen:wpaper:14-12} for an application to hierarchical firm data with both firm/employer and establishment level imputation). However, these differences may affect the observed density of the data within industry-year categories, and therefore the overall comparability. 

Finally, due to a feature of \SynLBD{} that we did not fully explore, synthesis of data in the last year of the data generally was of poor quality. For some industry-country pairs, this also happened in the first year. We dropped those observations. 

\subsection{Measuring outcomes}

In order to assess the outcomes of the experiment, we inspect analytical validity by various measures and also evaluate the extent of confidentiality protection. To check analytical validity, we compare basic univariate time series between the synthetic and confidential data (employment, entity entry and exit rates, job creation and destruction rates), and the distribution of entities (firms and establishment, depending on country),  employment, and payroll across time by industry. For a more complex assessment, we compute a dynamic panel data model of economic (employment) growth on each data set. 
We computed, but do not report here the confidence interval overlap measure (CIO) proposed by \citet{tas2006} in all these evaluations.\footnote{The full parameter estimates and the computed CIO are available in our replication materials \parencite{SIT-paper-repo}.}
The CIO is a popular measure when evaluating the validity for specific analyses. It evaluates how much the confidence intervals of the original data and the synthetic data overlap. We did not find this measure to be useful in our context. Most of our analyses are based on millions of records, and observed confidence intervals were so small that confidence intervals (almost) never overlap even when the estimates between the original data and the synthetic data are quite close.

To provide a more comprehensive measure of  quality of the synthetic data relative to the confidential data, we compute the $pMSE$ \parencite[propensity score mean-squared error,][]{Woo_Reiter_Oganian_Karr_2009,SnokeSlavkovic2018,Snoke_RSSA2018}: the mean-squared error of the predicted probabilities (i.e., propensity scores) for those two databases. Specifically, $pMSE$ is a metric to assess how well we are able to discern the high distributional similarity between synthetic data and confidential data. We follow  \textcite{Woo_Reiter_Oganian_Karr_2009} and \textcite{SnokeSlavkovic2018} to calculate the $pMSE$, using the following algorithm:  
\begin{enumerate}
    \item Append the $n_1$ rows of the confidential database $X$ to the $n_2$ rows of the synthetic database $X^s$ to create $X^{comb}$ with $N=n_1 + n_2$ rows, where both $X$ and $X^s$ are in the long format.
    \item Create a variable $I_{et}$ denoting membership of an observation for entity $e$, $e=1,\ldots,E$, at time point $t$, $t=1,\ldots,T$, in the component databases,  $I_{et}=\{1: X^{comb}_{et} \in X^s\}$. $I_{et}$ takes on values of $1$ for the synthetic database and $0$ for the confidential database. 
    \item Fit the following generalised linear model to predict $I$
    \begin{eqnarray}	
        P(I_{et}=1) & = &g^{-1}(\beta_0 + \beta_{1} Emp_{et} + \beta_{2} Pay_{et} + Age_{et}^{T}\beta_{3} + \lambda_t + \gamma_i), \label{pMSE}
     \end{eqnarray}
         where $Emp_{et}$ is  log employment  of entity $e$ in year $t$, $Pay_{et}$ is  log payroll of entity $e$ in year $t$, $Age_{et}$ is a vector of age classes of entity $e$ in year $t$, $\lambda_t$ is a year fixed effect, $\gamma_i$ is an time-invariant industry-specific effect for the industry classification $i$ of entity $e$, and $g$ is an appropriate link function (in this case, the logit link).
\item Calculate the predicted probabilities, $\hat{p}_{et}$.
    \item Compute  $pMSE=\frac{1}{N}\sum_{t=1}^T\sum_{e=1}^E(\hat{p}_{et} - c)^2$, where $c=n_2/N$.
\end{enumerate}
If $n_1 = n_2$, $pMSE$ = 0 means every $\hat{p}_{et}= 0.5$, and the two databases are distributionally indistinguishable, suggesting  high analytical validity. While the number of records in the synthetic data typically matches the number of records in the original data, i.e., $n_1 = n_2$, this does not necessarily hold in our application. Although the synthesis process ensures that the total number of entities is the same in both data sets, the years in which the entities are observed will generally differ between the original data and the synthetic data and thus the number of records in the long format will not necessarily match between the two data sets. For this reason we follow 
\citet{Woo_Reiter_Oganian_Karr_2009} and \citet{Snoke_RSSA2018} and use $c=n_2/N$ instead of fixing $c$ at 0.5. Using this more general definition, $c$ will always be the mean of the predicted propensity scores so that the $pMSE$ measures the average of the squared deviations from the mean, as intended. 

Since the $pMSE$ depends on the number of predictors included in the propensity score model, \textcite{Snoke_RSSA2018} derived the expected value and standard deviation for the $pMSE$ under the null hypothesis ($pMSE_0$) that the synthesis model is correct, i.e., it matches the true data generating process \parencite[Equation 1]{Snoke_RSSA2018}:
$$
E[pMSE_0] = (k-1)(1-c)^2 \frac{c}{N}
$$
and
$$
StDev[pMSE_0] = \sqrt{2(k-1)}(1-c)^2 \frac{c}{N}
$$
where $k$ is the number of synthesized variables used in the propensity model. To measure the analytical validity of the synthetic data, they suggest looking at the \textit{pMSE  ratio}
$$
pMSE ratio = \frac{\widehat{pMSE}}{E[pMSE_0]}
$$
and the \textit{standardized pMSE}
$$
pMSE_s =\frac{\widehat{pMSE}-E[pMSE_0]}{StDev[pMSE_0]},
$$
where $\widehat{pMSE}$ is the estimated pMSE based on the data at hand. Under the null hypothesis, the $pMSE$ ratio has an expectation of 1 and the expectation of the standardized $pMSE_s$ is zero.

\section{Analytical validity}
\label{sec:analytic}

\newcommand{\CanTableNote}{$LEAP$ is the Longitudinal Employment Analysis Program and $CanSynLBD$ is the Canadian synthetic database based on LEAP. }

In the following figures, the results for the Canadian data are shown in the left panels, and the German data in the right panels. In all cases, the Canadian data are reported for the entire private sector,  including the manufacturing sector but excluding  the public sector industries (NAICS 61, 62, and 91). German results are for two WZ2003 industries.

\subsection{Entity Characteristics}

\begin{figure}[t]
  \begin{subfigure}[h]{0.48\linewidth}
    \includegraphics[trim=0 40 0 0,clip, width=\linewidth]{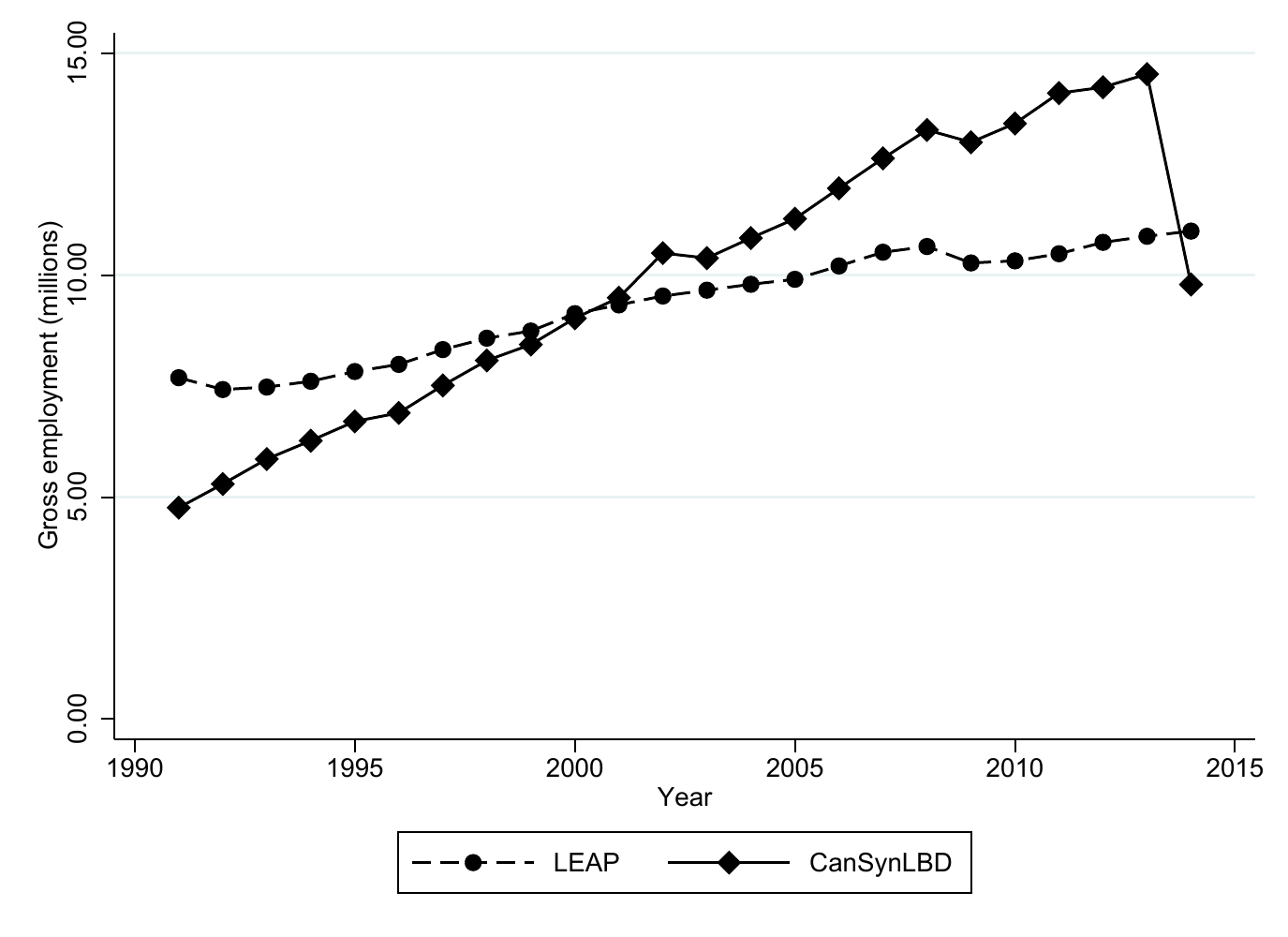}
  \end{subfigure}
\hfill
  \begin{subfigure}[h]{0.48\linewidth}
     \includegraphics[trim=0 40 0 0,clip,width=\linewidth]{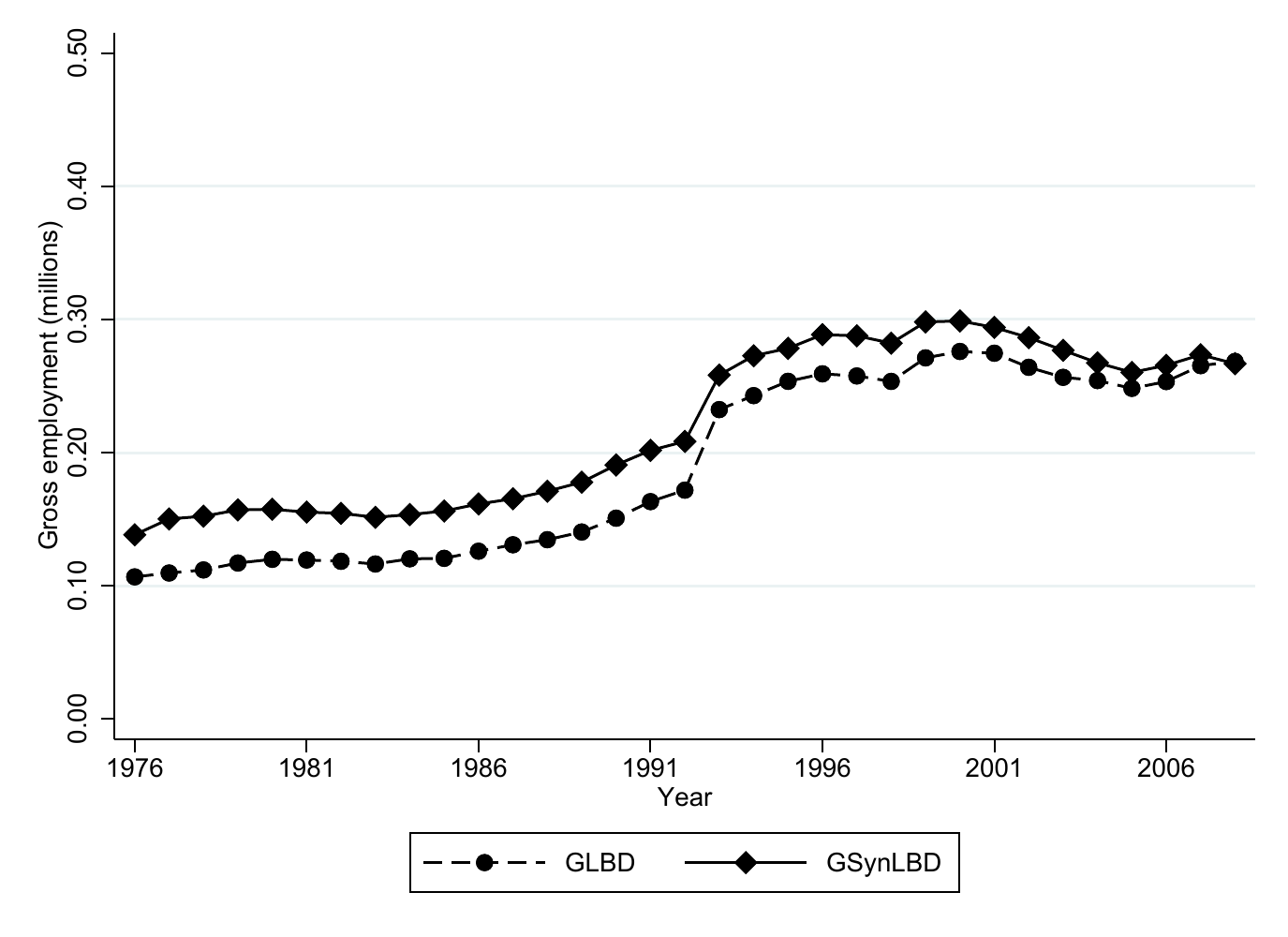}
  \end{subfigure}\\
  \begin{subfigure}[h]{0.48\linewidth}
    \includegraphics[trim=0 0 0 -20,clip,width=\linewidth]{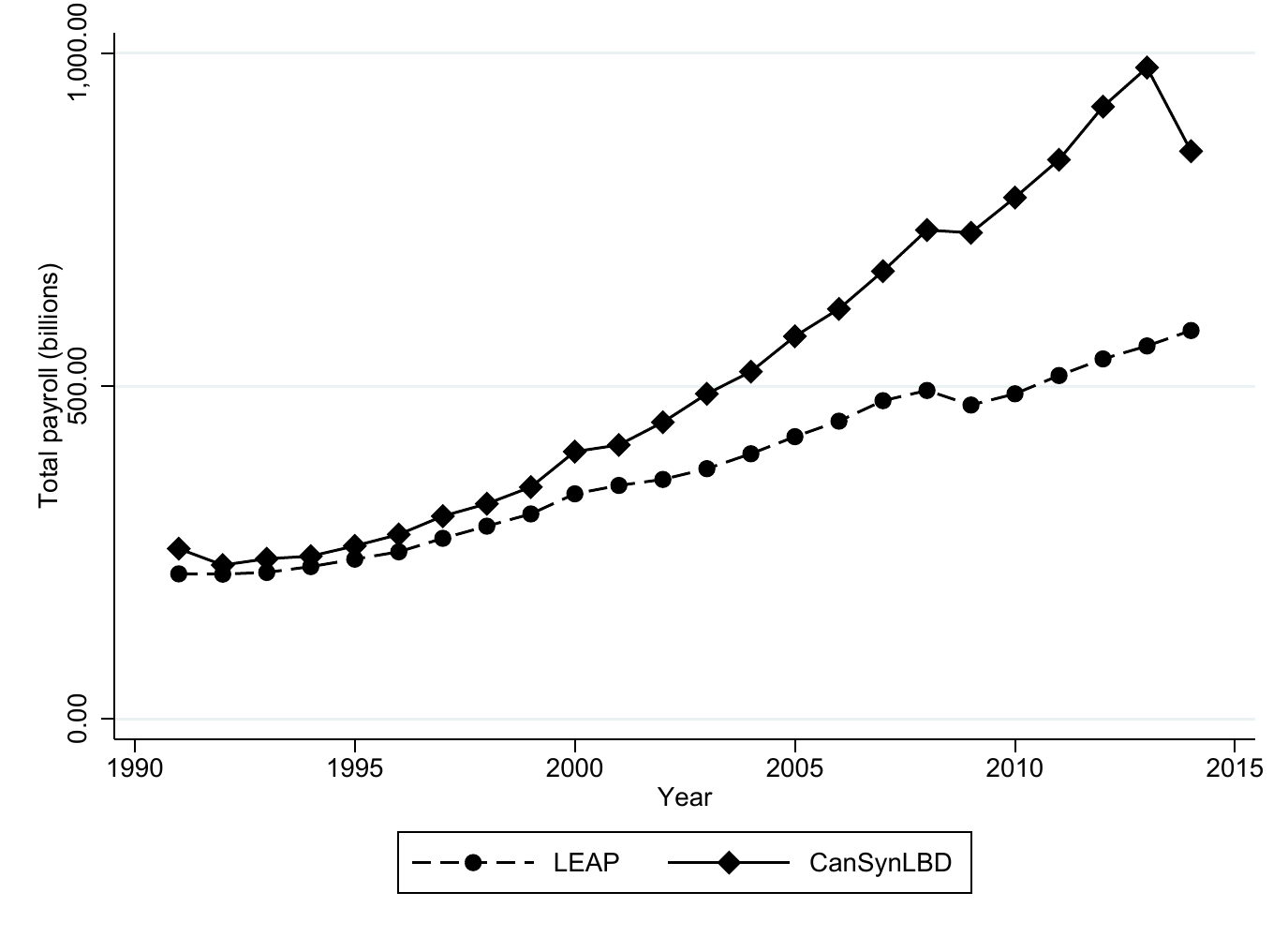}
   \caption{CanSynLBD}
   \end{subfigure}
\hfill
   \begin{subfigure}[h]{0.48\linewidth}
     \includegraphics[trim=0 0 0 -20,clip,width=\linewidth]{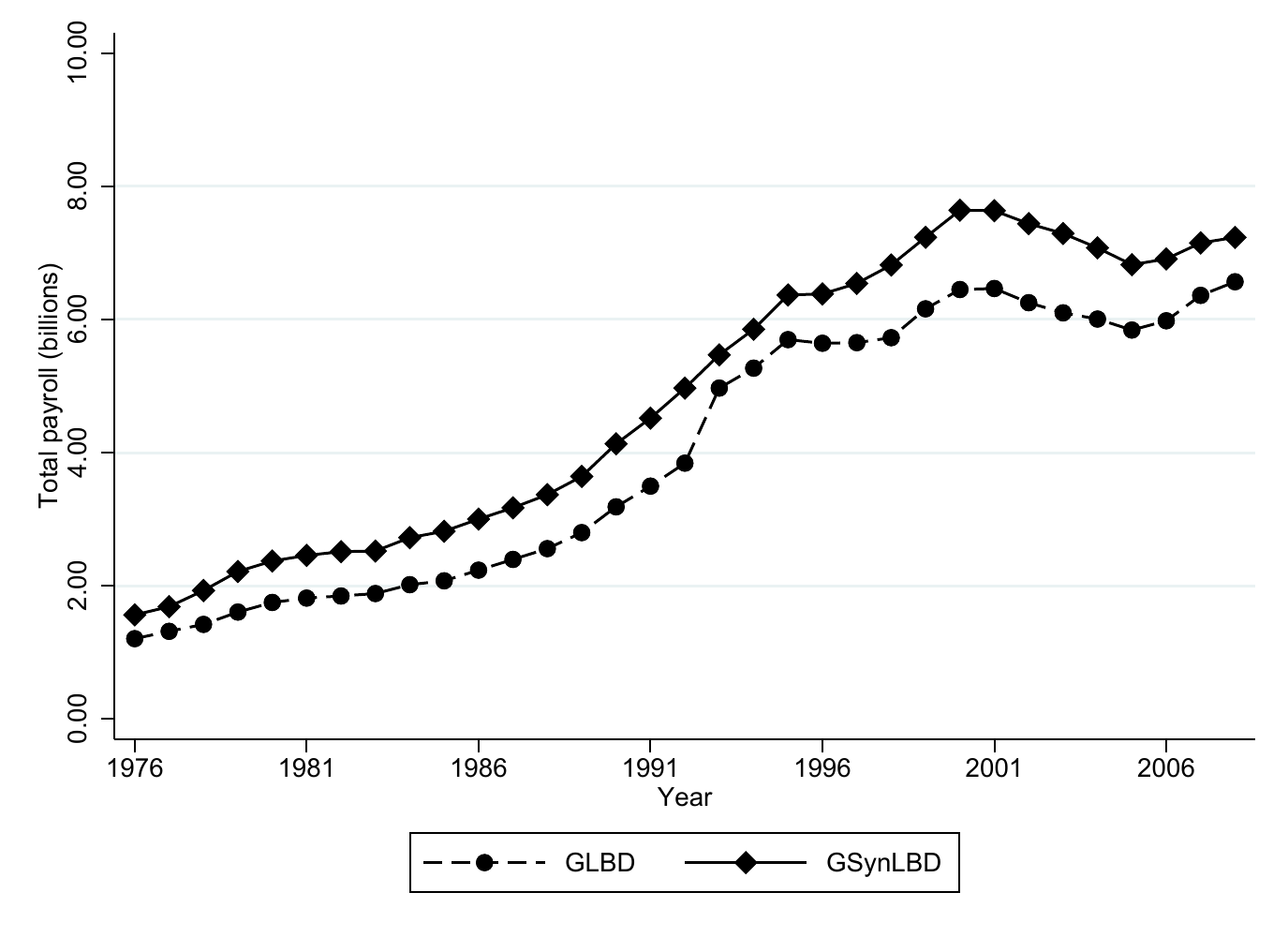}
     \caption{GSynLBD}
   \end{subfigure}\caption{Gross employment level (upper panels) and total payroll (lower panels) by year.}\label{fig:entity_chracteristics}
\end{figure}

Figure \ref{fig:entity_chracteristics} shows a comparison between the synthetic data and the original data for gross employment level (upper panels) and total payroll (lower panels) by year. While the general trends are preserved for both data sources, the results for the German synthetic data resemble the trends from the original data more closely. For the Canadian data the positive trends over time are generally overestimated. However, in both cases, levels are mostly overestimated. These patterns are not robust. When considering the   manufacturing sector in Canada (Figure 8 
in the  Appendix), trends are better matched, but a significant \textit{negative} bias is present in levels.

\subsection{Dynamics of Job Flows}

\begin{figure}[t]
\begin{subfigure}[h]{0.48\linewidth}
\includegraphics[trim=0 40 0 0,clip, width=\linewidth]{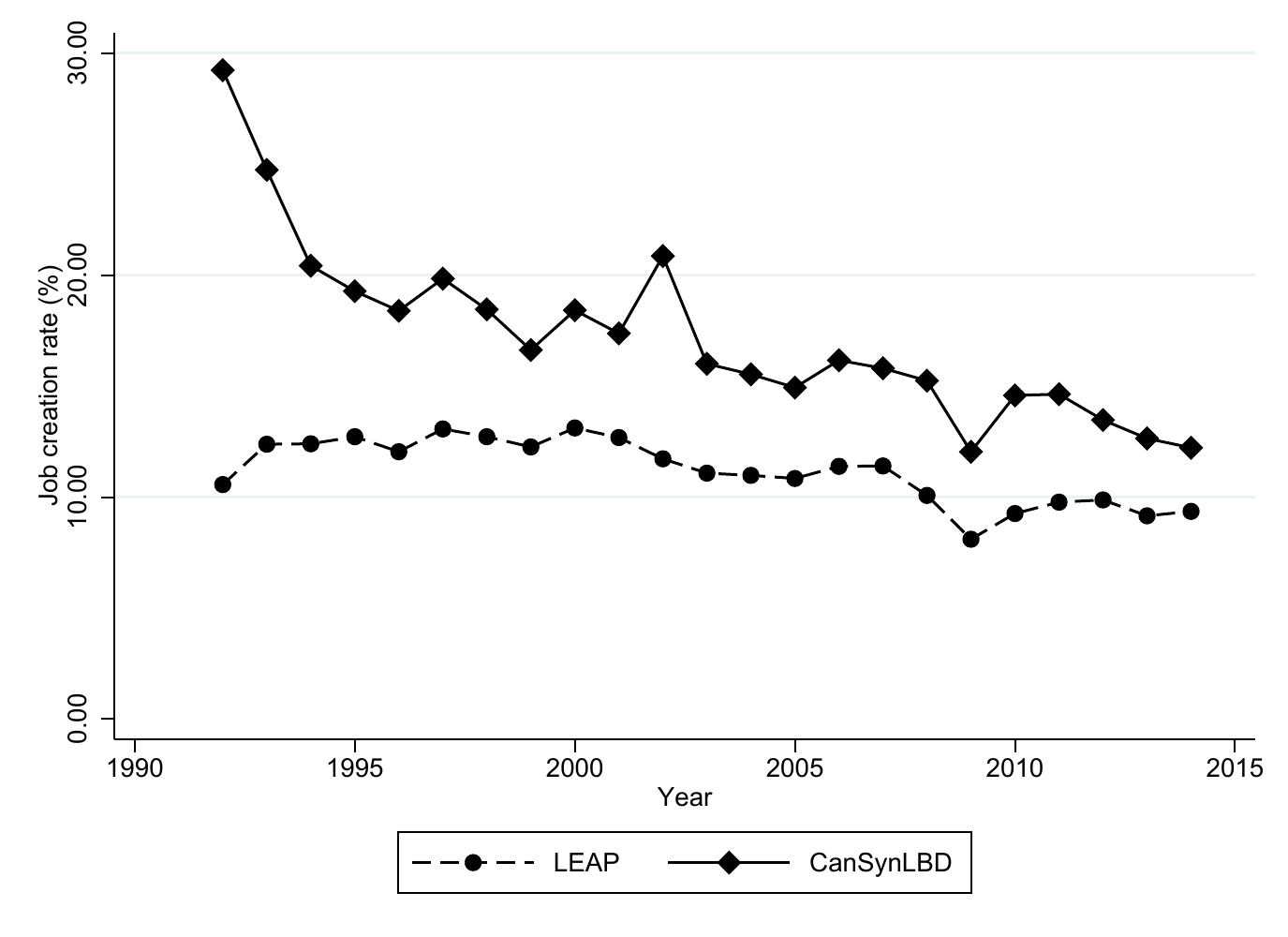}
\end{subfigure}
\hfill
\begin{subfigure}[h]{0.48\linewidth}
\includegraphics[trim=0 40 0 0,clip,width=\linewidth]{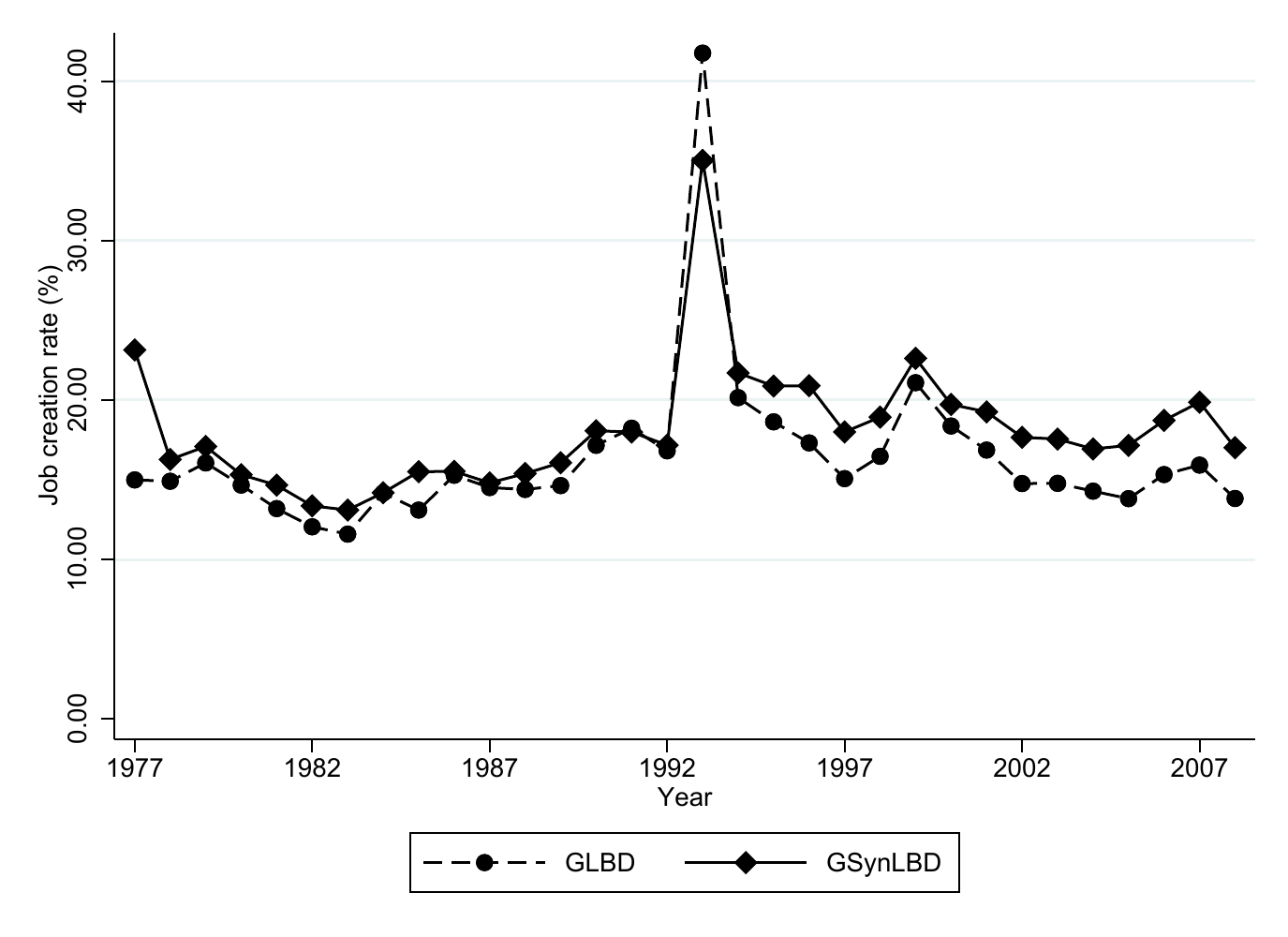}
\end{subfigure}\\
\begin{subfigure}[h]{0.48\linewidth}
\includegraphics[trim=0 0 0 -20,clip,width=\linewidth]{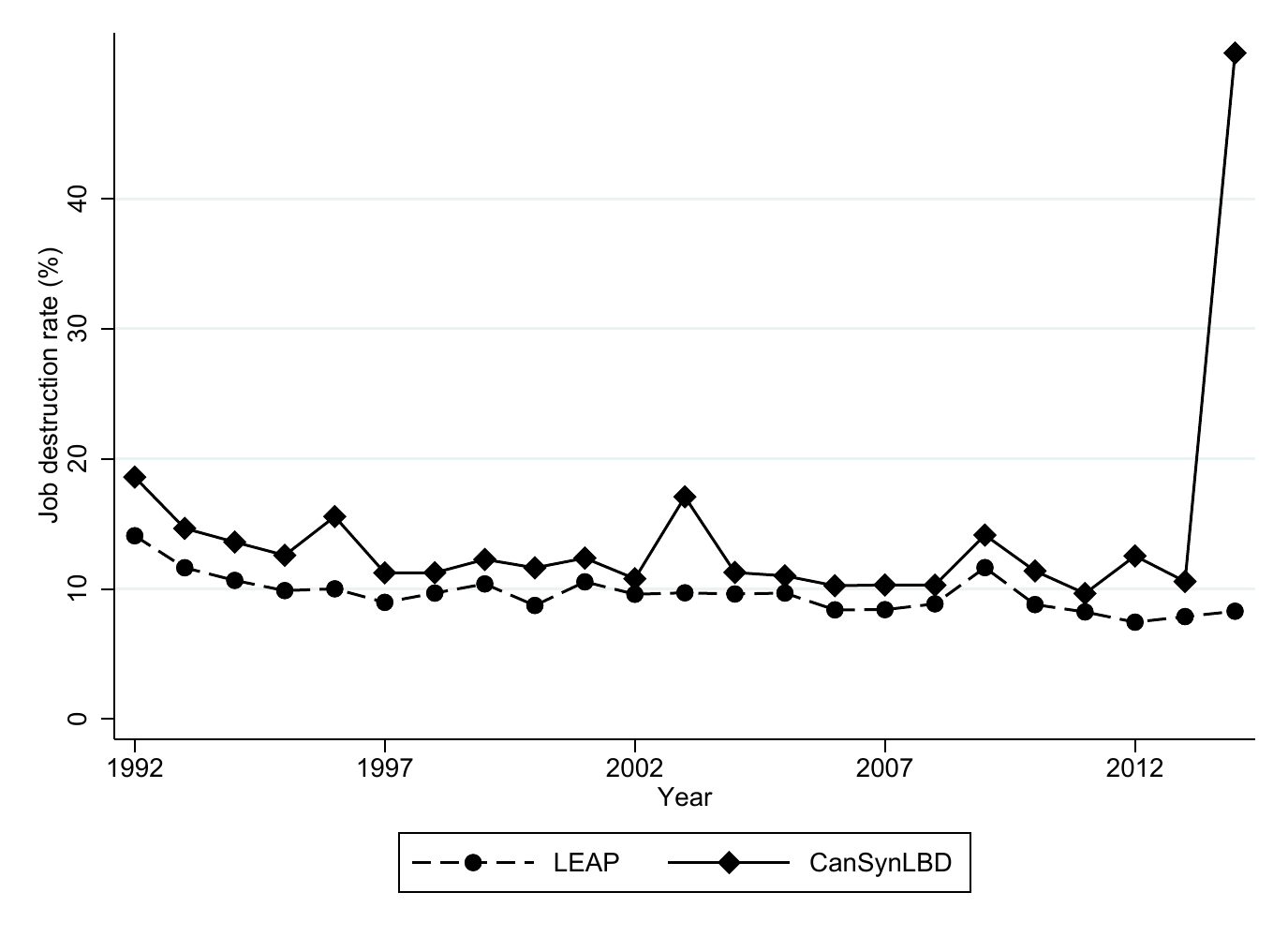}
\caption{CanSynLBD}
\end{subfigure}
\hfill
\begin{subfigure}[h]{0.48\linewidth}
\includegraphics[trim=0 0 0 -20,clip,width=\linewidth]{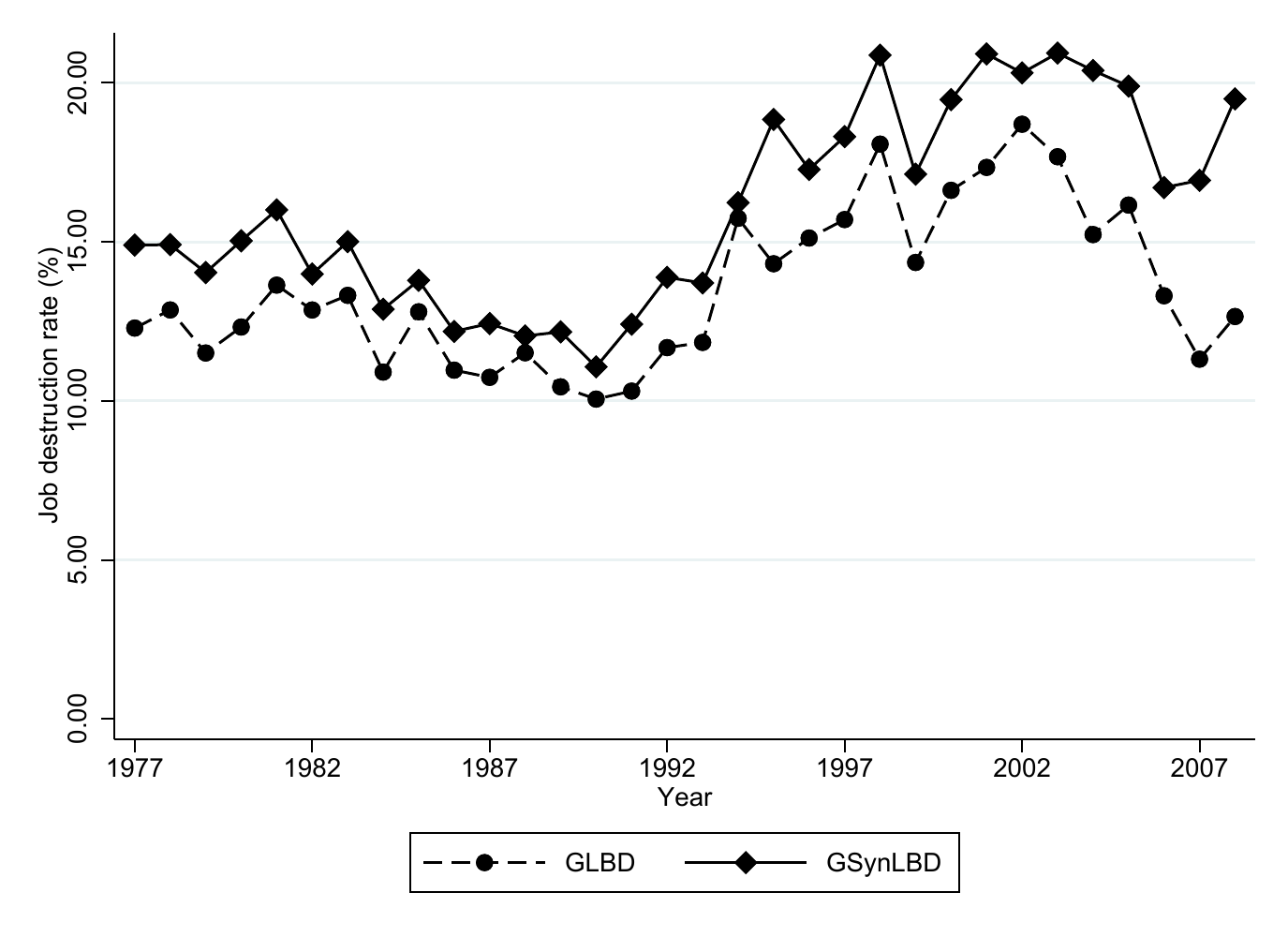}
\caption{GSynLBD}
\end{subfigure}\caption{Job creation rates (upper panels) and job destruction rates (lower panels) by year.}\label{fig:job_flows}
\end{figure}
 
Key statistics commonly computed from business registers such as the LEAP or the BHP include job flows over time. Following \citet{DavisHaltiwangerSchuh}, job creation is defined as the sum of all employment gains from expanding firms from year $t-1$ to year $t$ including entry firms. The job destruction rate is defined as the sum of all employment losses from contracting firms from year $t-1$ to year $t$ including exiting firms. Figure~\ref{fig:job_flows} depicts  job creation rates (upper panels) and destruction rates (lower panels). The general levels and trends are preserved for both data sources, but the time-series align more closely for the German data. Even the substantial increase in job creations in 1993, which can be attributed to the integration of the data from Eastern Germany after reunification, is remarkably well preserved in the synthetic data. Still, there seems to be a small but systematic overestimation of job creation and destruction rates in both synthetic data sources. The substantial deviation in the job destruction rate in the last year of CanSynLBD is an artefact  requiring further investigation.\footnote{The results for the Canadian manufacturing sector are included in Figure 9
in the  Appendix, and are comparable to the results for the entire private sector.}

\subsection{Entity Dynamics}

\begin{figure}[ht]
   \begin{subfigure}[h]{0.48\linewidth}
     \includegraphics[trim=0 40 0 0,clip, width=\linewidth]{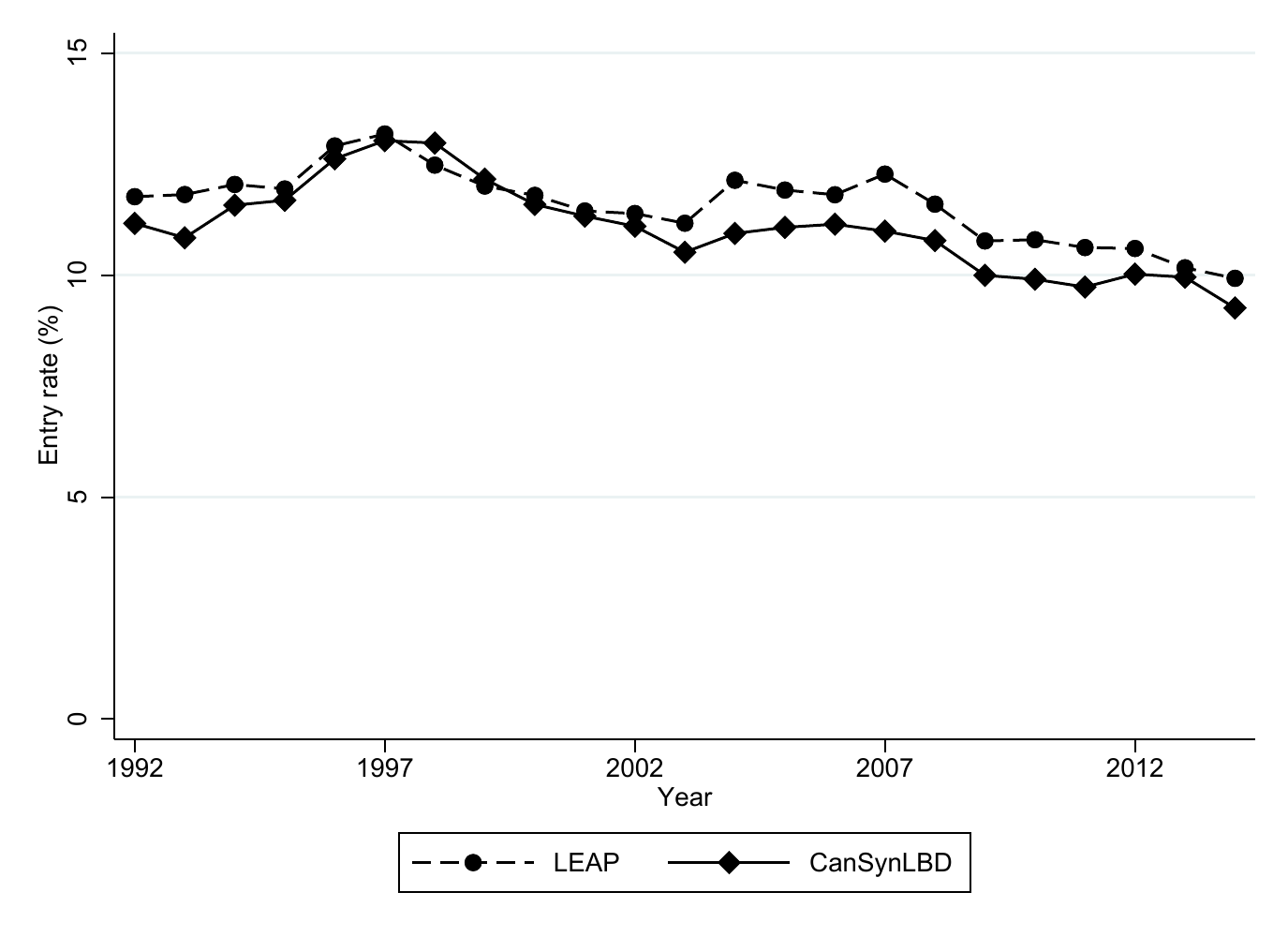}
   \end{subfigure}
\hfill
   \begin{subfigure}[h]{0.48\linewidth}
      \includegraphics[trim=0 40 0 0,clip,width=\linewidth]{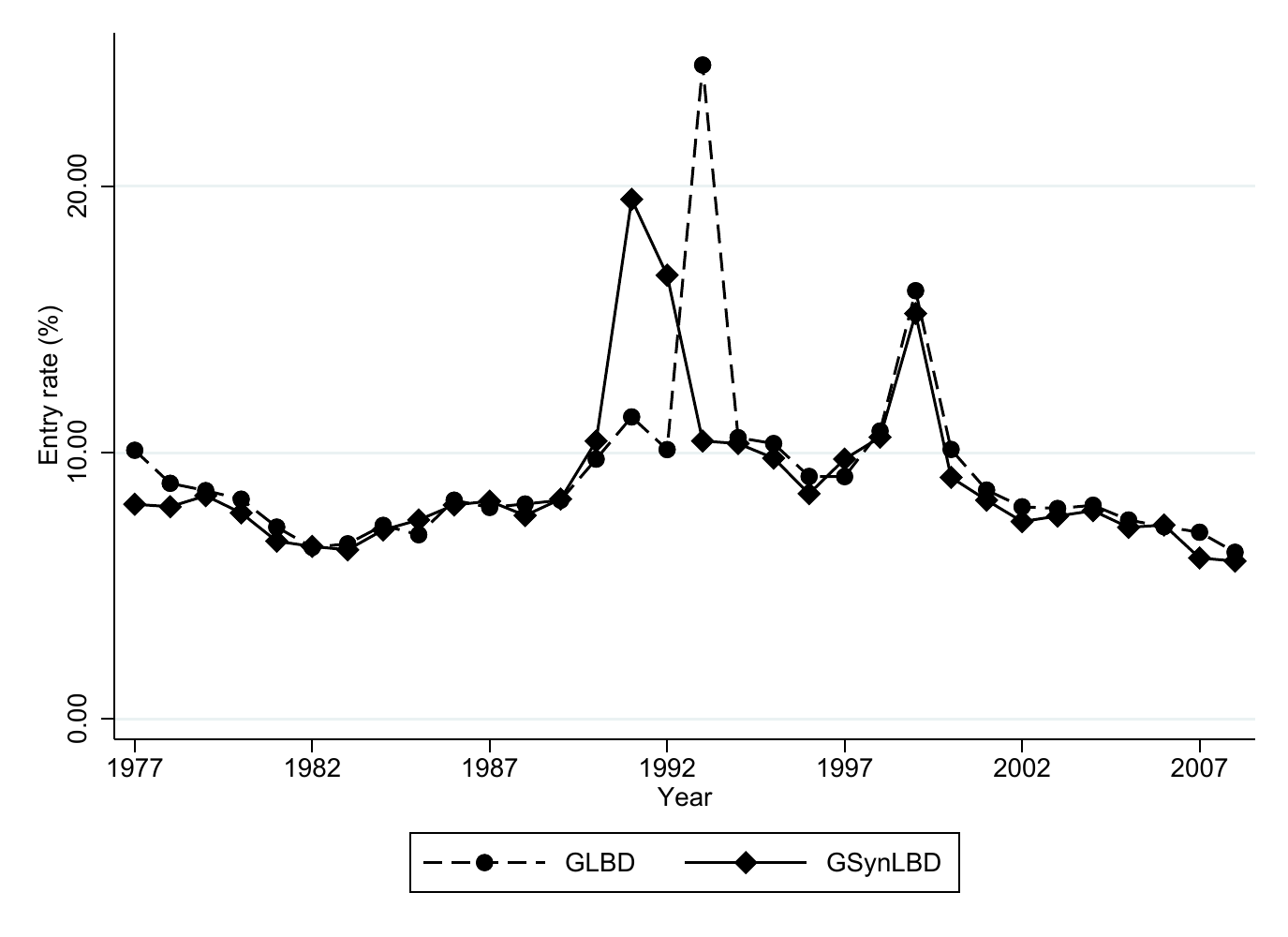}
   \end{subfigure}\\
   \begin{subfigure}[h]{0.48\linewidth}
     \includegraphics[trim=0 0 0 -20,clip,width=\linewidth]{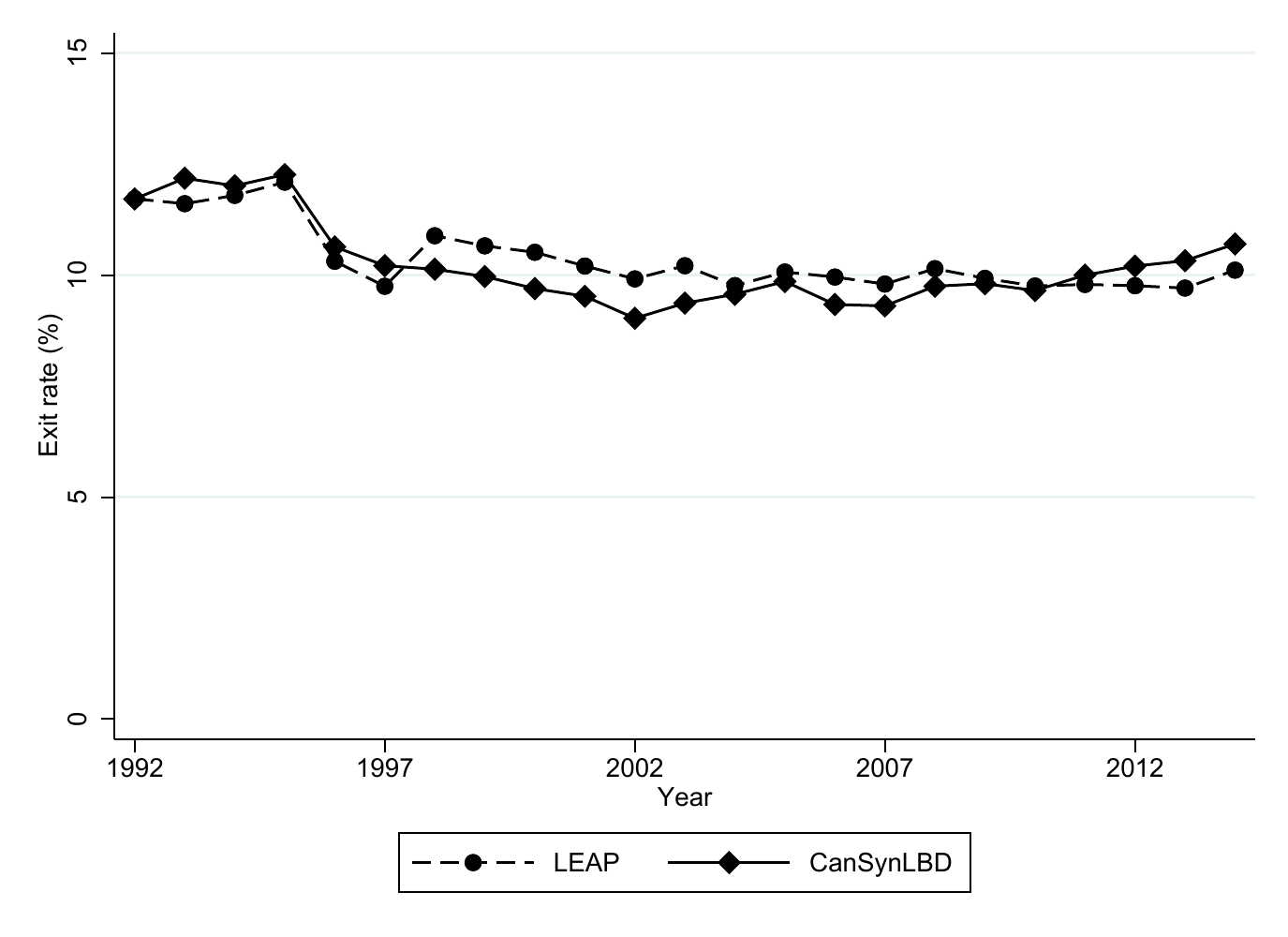}
    \caption{CanSynLBD}
   \end{subfigure}
\hfill
   \begin{subfigure}[h]{0.48\linewidth}
     \includegraphics[trim=0 0 0 -20,clip,width=\linewidth]{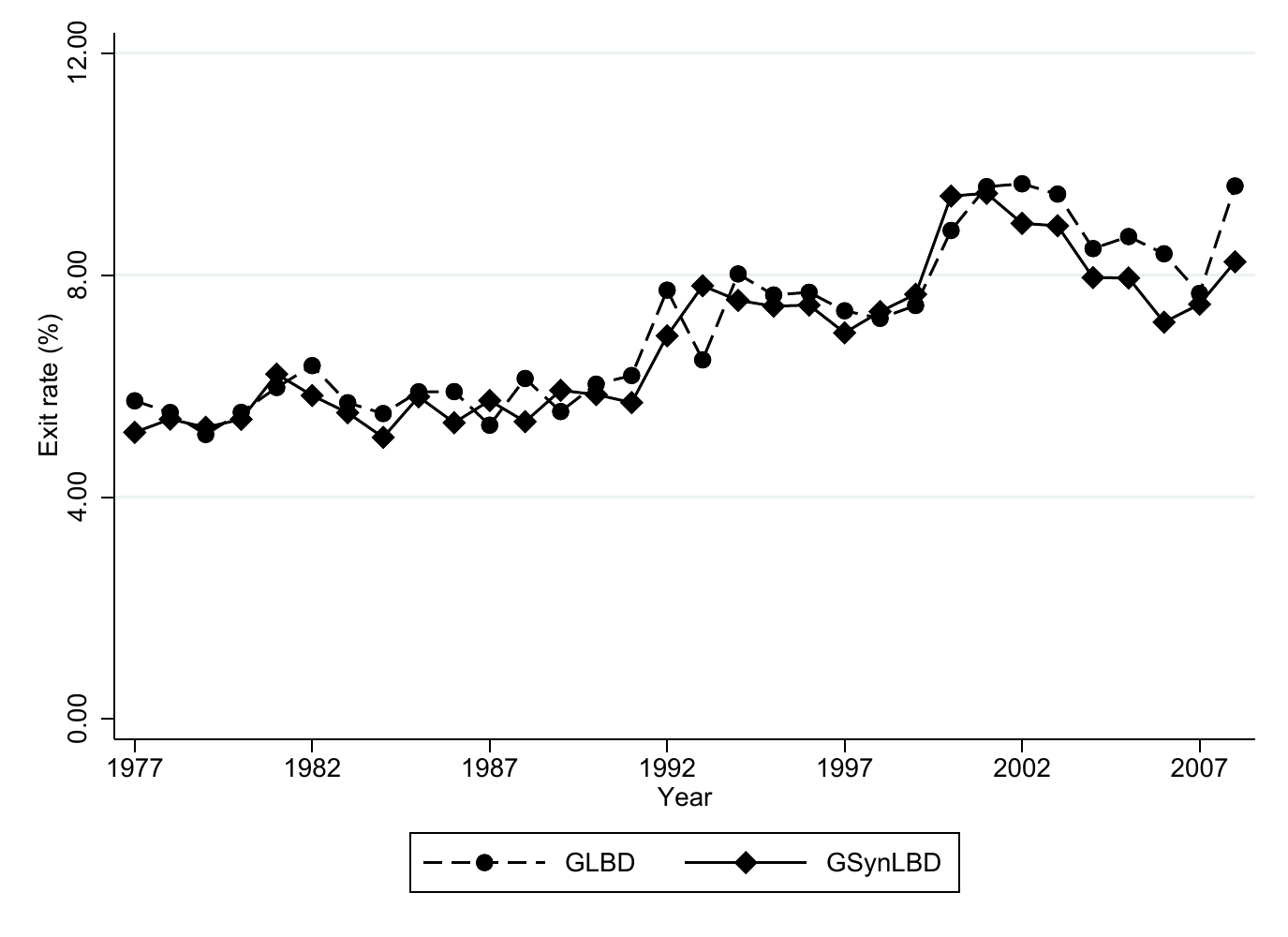}
     \caption{GSynLBD}
   \end{subfigure}
\caption{Entry rates (upper panels) and exit rates (lower panels) by year.\label{fig:FirmDynamics}}
\end{figure}

To assess how well the synthetic data capture entity dynamics, we also compute entry and exit rates, i.e. how many new entities appear in the data and how many cease to exist relative to the population of entities in a specific year.\footnote{As described in Section~\ref{sec:data}, for both countries' data, corrections based on worker flows have been applied, correcting for any bias due to legal reconfiguration of economic entities.} Figure~\ref{fig:FirmDynamics} shows that those rates are very well preserved for both data sources. 

Only the (delayed) re-unification spike in the entry rates in the German data is not preserved correctly. The confidential data show a large spike in entry rates in 1993. In that year, detailed information about Eastern German establishments was integrated for the first time. However, the synthetic data shows increased entry rates in the two previous years. We speculate that this occurs due to incomplete data in the confidential data: Establishments were successively integrated into the data starting in 1991, but  many East German establishments did not report payroll and number of employees  in the first two years. Thus, records existed in the original data, but the establishment size is reported as missing. Such a combination is not possible in the synthetic data. The synthesis models are constructed to ensure that whenever an establishment exists, it has to have a positive number of employees. Since entry rates are computed by looking at whether the employment information changed from missing to a positive value, most of the Eastern German establishments only exist from 1993 on-wards in the original data, but from 1991 in the synthetic data.

The second, smaller spike in the entry rate in the German data occurs in 1999. In that year, employers were required to report  marginally employed workers  for the first time. Some establishments exclusively employ marginally employed workers, and will thus appear for the first time in the data after 1999. The synthetic data preserves this pattern.

\subsection{Distribution of variables across time and industry}

\begin{figure}[!ht]
\begin{subfigure}[h]{0.48\linewidth}
\includegraphics[trim=0 10 0 0,clip, width=\linewidth]{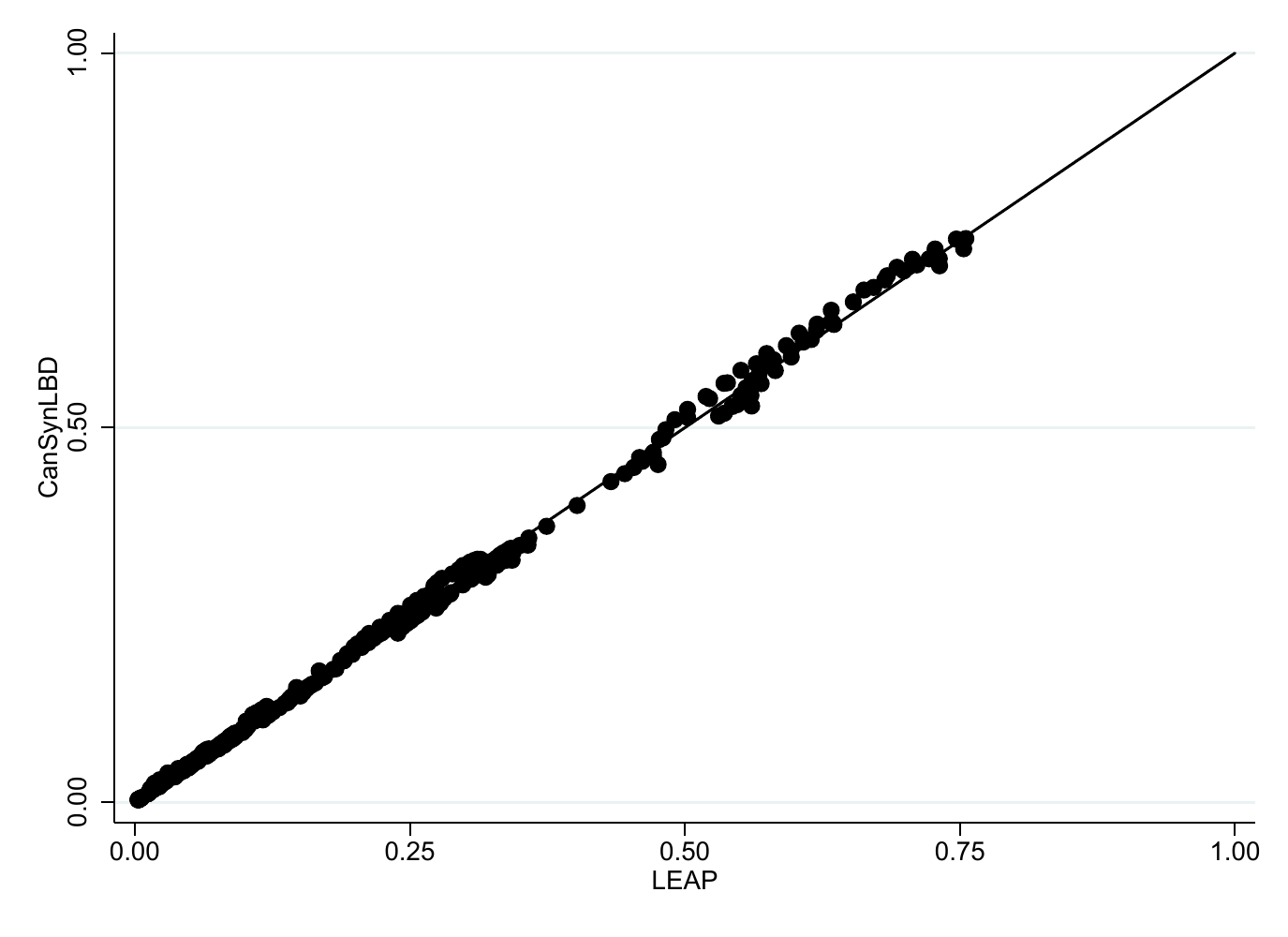}
\end{subfigure}[t]
\hfill
\begin{subfigure}[h]{0.48\linewidth}
\includegraphics[trim=0 10 0 0,clip,width=\linewidth]{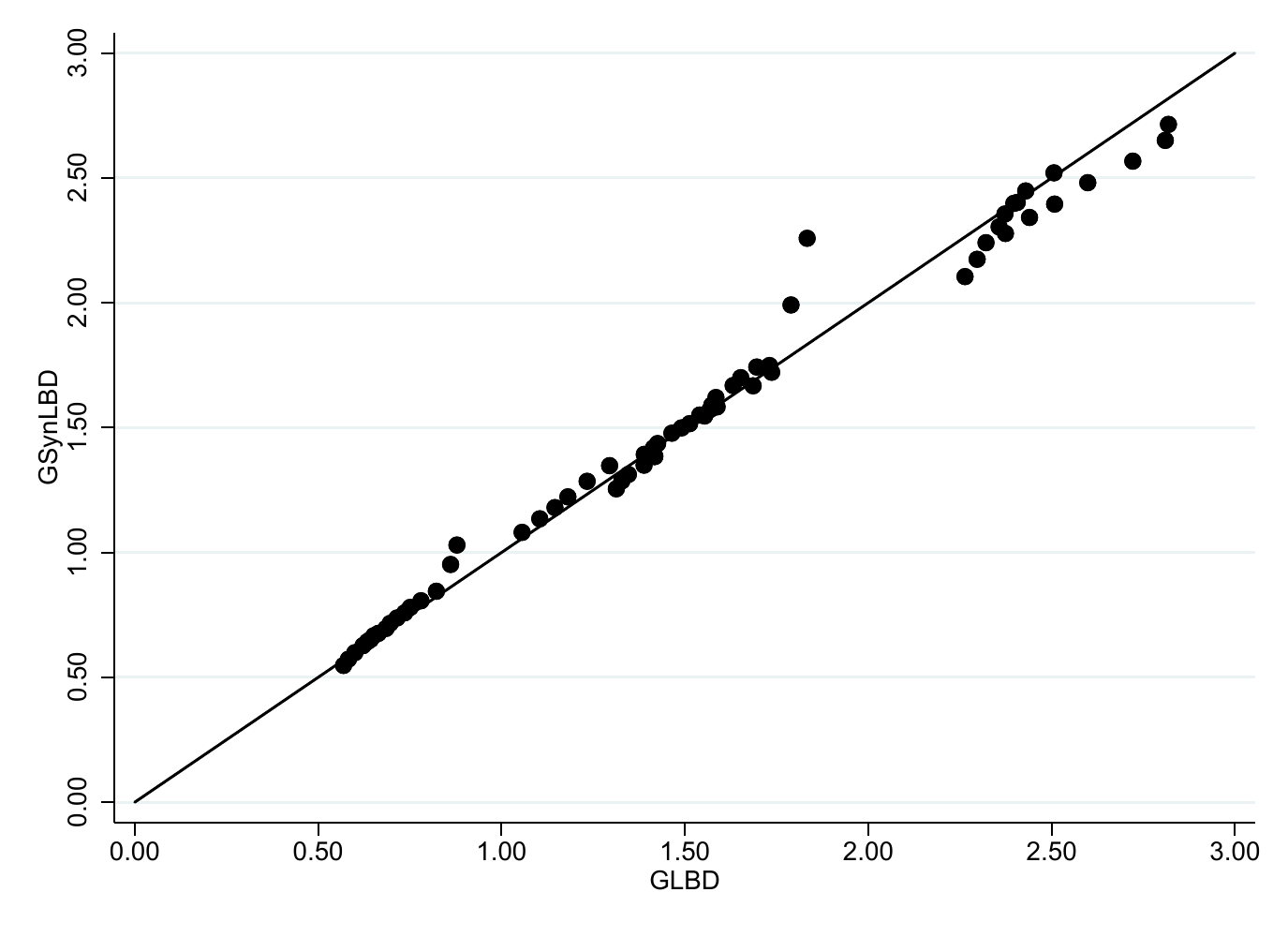}
\end{subfigure}\\
\begin{subfigure}[h]{0.48\linewidth}
\includegraphics[trim=0 10 0 -20,clip,width=\linewidth]{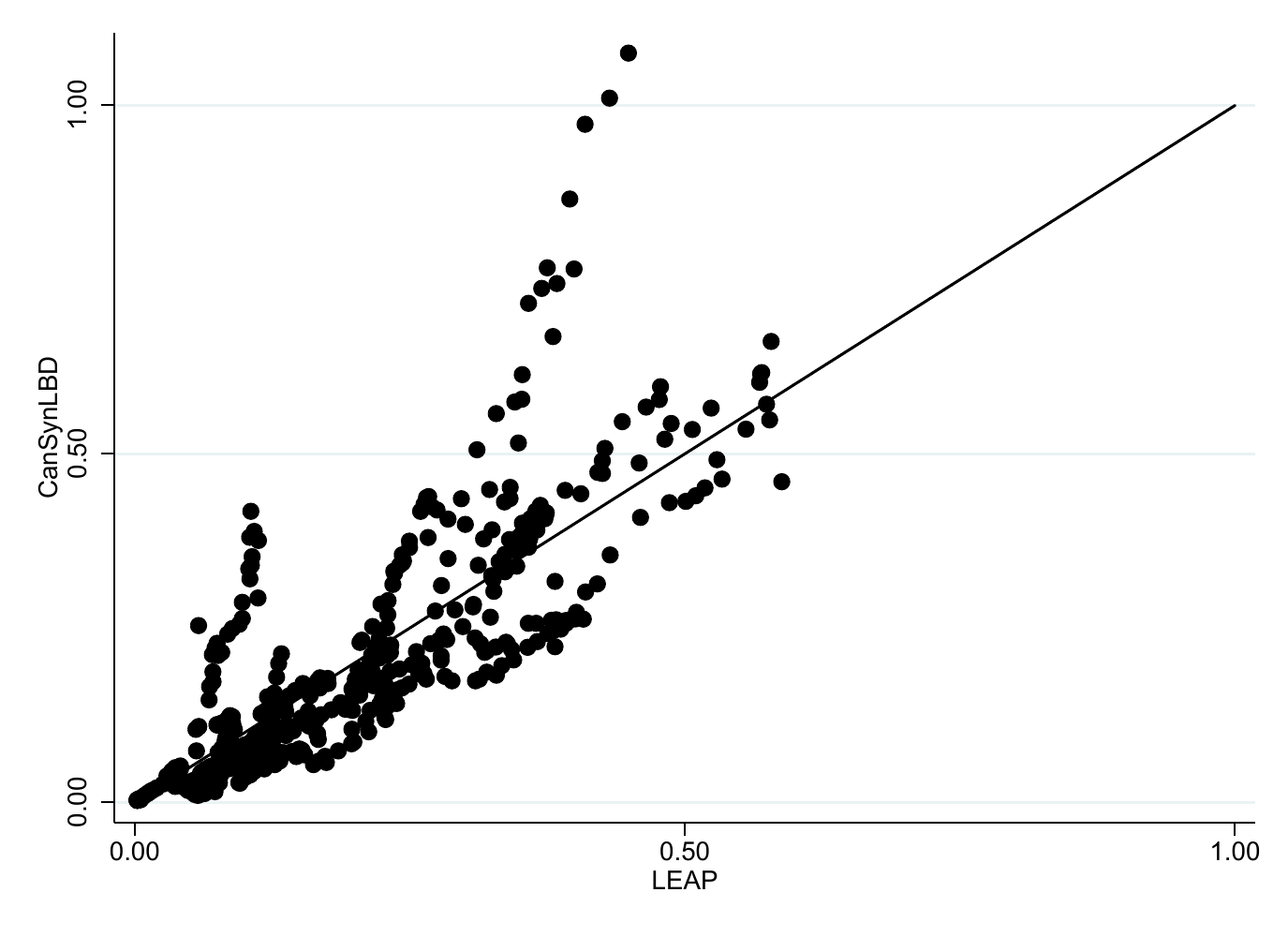}
\end{subfigure}
\hfill
\begin{subfigure}[h]{0.48\linewidth}
\includegraphics[trim=0 10 0 -20,clip,width=\linewidth]{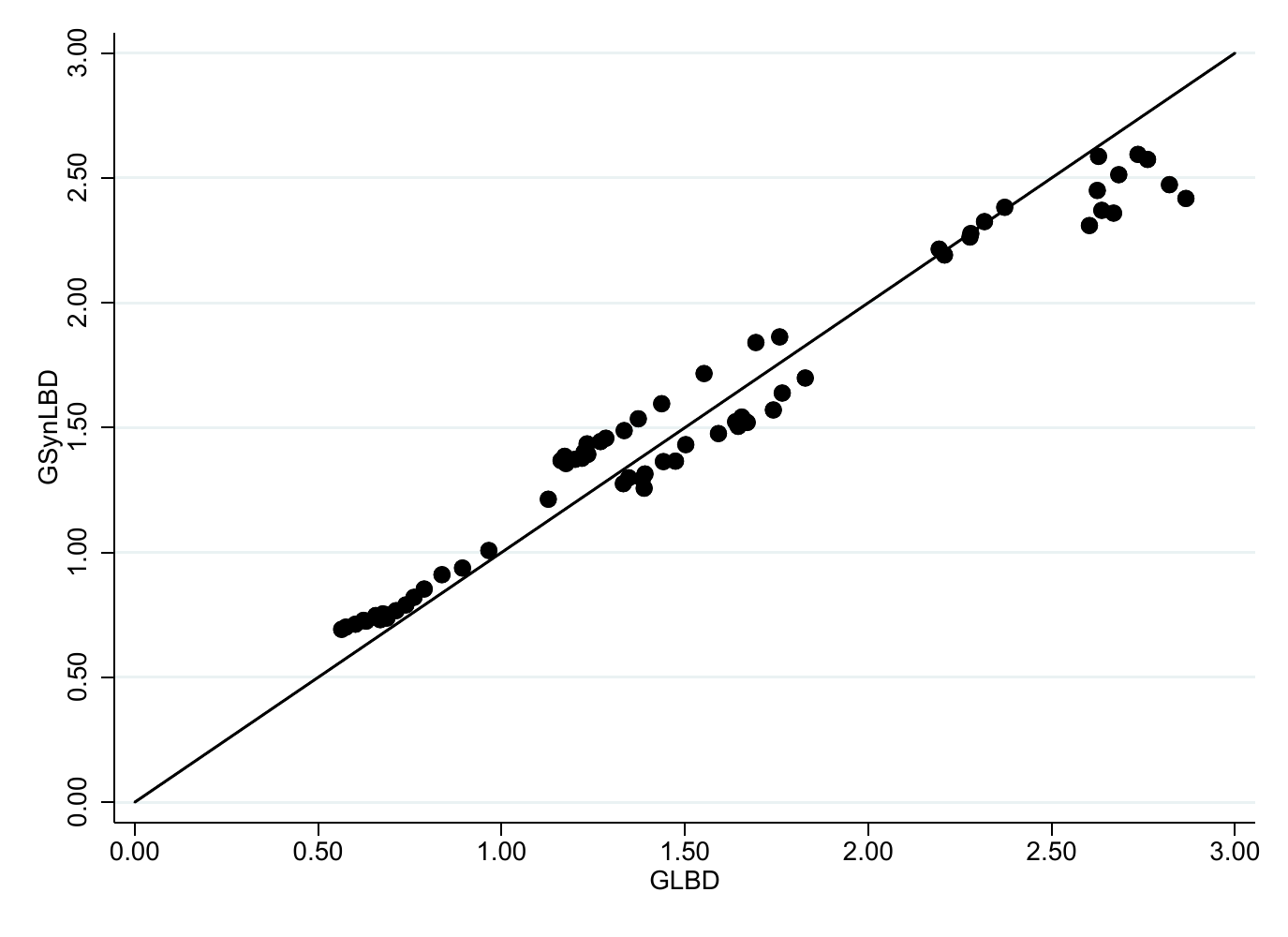}
\end{subfigure}\\
\begin{subfigure}[h]{0.48\linewidth}
\includegraphics[trim=0 0 0 -20,clip,width=\linewidth]{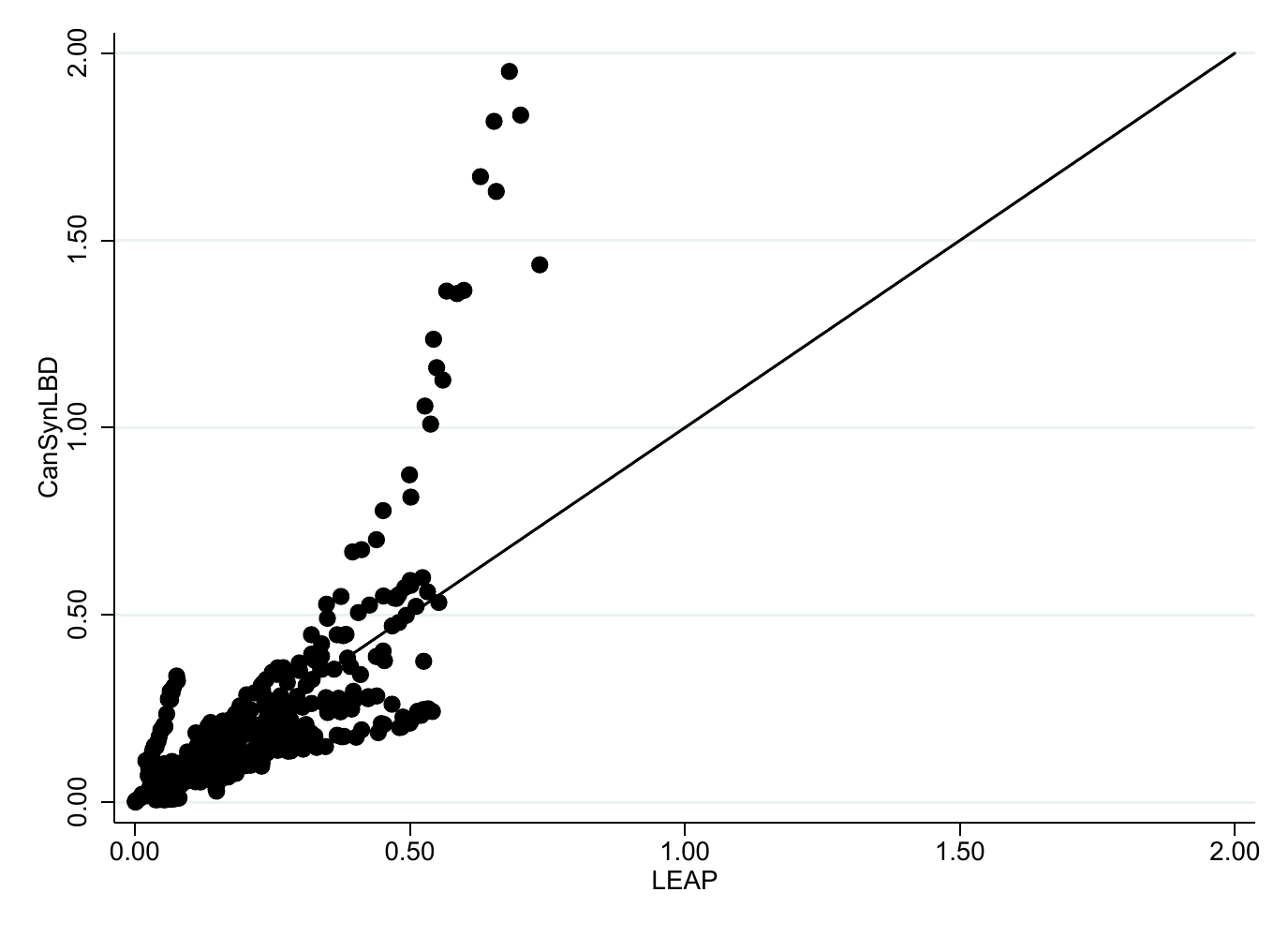}
\caption{CanSynLBD}
\end{subfigure}
\hfill
\begin{subfigure}[h]{0.48\linewidth}
\includegraphics[trim=0 0 0 -20,clip,width=\linewidth]{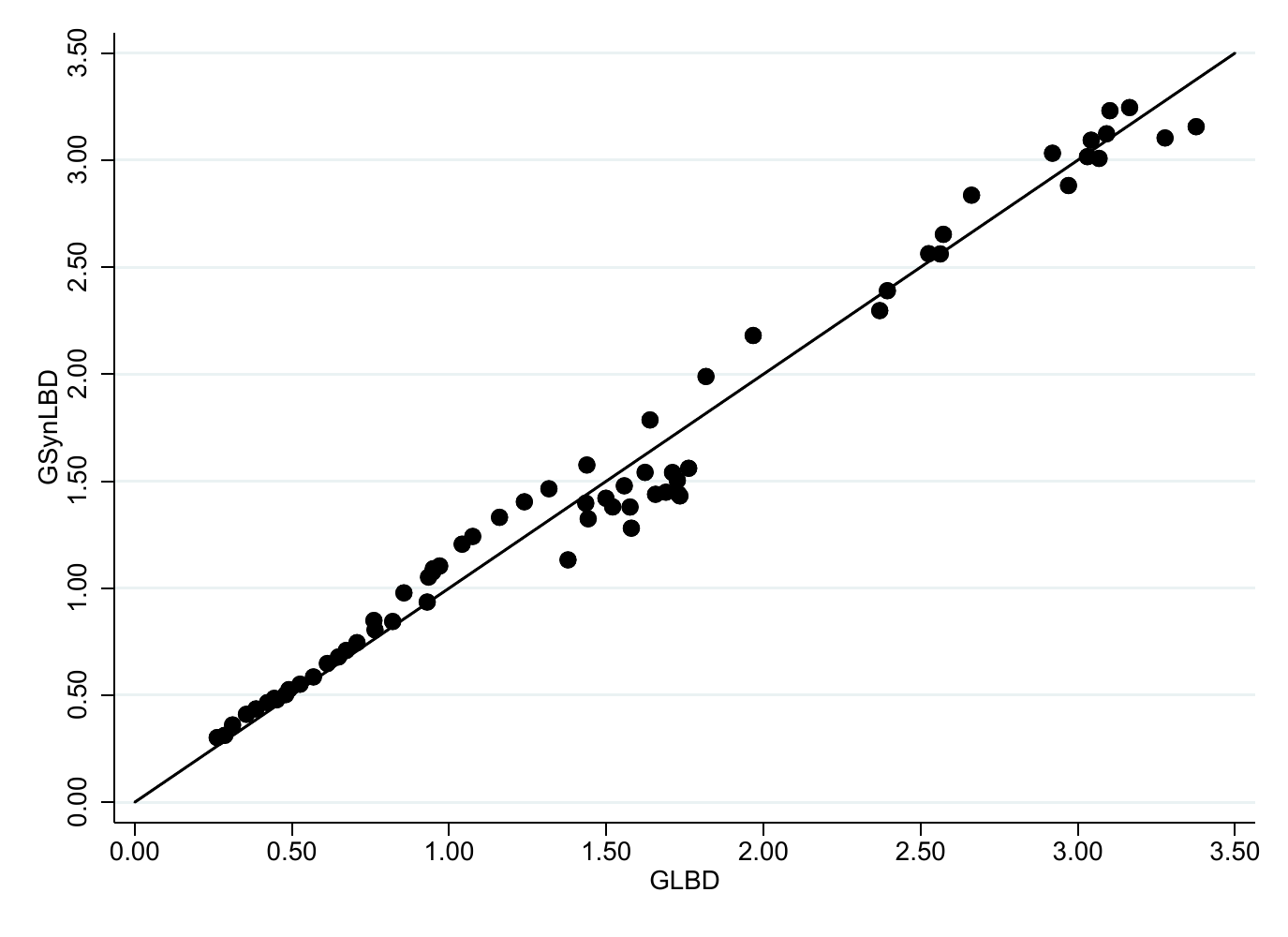}
\caption{GSynLBD}
\end{subfigure}\caption{Share of entities (upper panels), share of employment (middle panels), and share of payroll (lower panels) by year and industry.}\label{fig:FirmShare}
\end{figure}

The \SynLBD{} code  ensures that the total number of entities that ever exist within the considered time frame  matches exactly between the original data and the synthetic data. But  each entity's entry and exit date are synthesized, and the total number of entities at any particular point in time may differ, and with it employment and payroll. To investigate how well the information is preserved at any given point in time, we compute the following statistic:
\begin{equation}
    \label{eq:share_employment}
x_{its} = X_{its}/\sum_{i} \sum_{t} X_{its}, 
\end{equation}
where $i$ is the index for the industry (aggregated to the two digit level for the Canadian data), $t$ is the index for the year and $s$ denotes the data source (original or synthetic). $X_{its}=\sum_j X_{itsj}$, $j=1,\ldots,n_{its}$ is the variable of interest aggregated at the industry level and $n_{its}$ is the number of entities in industry $i$ at time point $t$ in data source $s$. 
To compute the statistic provided in Equation (\ref{eq:share_employment}), this number is then divided by the total of the variable of interest aggregated across all industries and years.
Figure~\ref{fig:FirmShare} plots the results from the original data against the results from the synthetic data for the  number of entities, employment, and {payroll}. If the information is well preserved, all points should be close to the 45 degree line.

We find that the share of entities is well preserved for both data sources, but share of employment and share of payroll vary more in the Canadian data with an upward bias for the larger shares. 
It should be noted that the German data shown here and elsewhere in this paper only contain data from two industries, whereas the Canadian data contains nearly all  available industry codes at the two digit level. Thus, results from Canada are expected to be more diverse. 
When only considering the Canadian manufacturing sector (see Figure 10 
in the  Appendix), less bias is present.

\subsection{Modelling strategy}

To assess how well the synthetic data perform in a more complex model and in the context of an analyst's modelling strategy, we simulate how a macroeconomist (the typical user of these data) might approach the problem of estimating a model for the evolution of employment if only the synthetic data are available. The analyst will consider both the literature and the data to propose a meaningful model. In doing so, a sequence of models will be proposed, and tests or theory brought to bear on their merits, potentially rejecting their appropriateness. In doing so, the outcome that the analyst obtains from following that strategy using the synthetic data should not diverge substantially from the outcome they would obtain when using the (inaccessible) confidential data. The specific parameter estimates obtained, and the actual model retained, are not the goal of this exercise --- the focus is on the process.

\begin{figure}
\centering 
\tikzset{
    fontscale/.style = {font=\relsize{#1}},
>=stealth',
punkt/.style={
           rectangle,
           rounded corners,
           draw=black, very thick,
           text width=9em,
           minimum height=2em,
           fontscale=0.8,
           text centered},
pil/.style={
           ->,
           thick,
           shorten <=2pt,
           shorten >=2pt,}
}
\tikzstyle{decision} = [diamond, draw, 
    text width=4.5em, text badly centered, node distance=3cm, inner sep=0pt]
\tikzstyle{line} = [draw, -latex']

\resizebox{\columnwidth}{!}{

\begin{tikzpicture}[node distance=1cm, auto,]
\node[punkt] (ols) {OLS};
 \node[right=0.1cm of ols] (m1) {};
 \node[punkt, inner sep=5pt,right=0.1cm of m1]
 (gmm) {GMM};
 \node[right=0.1cm of gmm] (m2) {};
 \node[punkt, inner sep=5pt,right=0.1cm of m2]
 (sgmm) {System GMM};
 \node[right=0.1cm of sgmm] (m3) {};
 \node[punkt, inner sep=5pt,right=0.1cm of m3]
 (sgmmma) {System GMM MA};
 \node[above of=m2] (m4) {};
 \node[above of= m4]
 (analyst) {Analyst}
 (analyst.east) edge[pil, bend left=22] node {\textit{specifies}} (sgmmma)
 (analyst.west) edge[pil, bend right=22] node {\textit{specifies}} (ols)
 (analyst) edge[pil, bend right=10]  (gmm)
 (analyst) edge[pil, bend left=10]  (sgmm);
 \node[decision,below=0.5cm of m1] (test1) {Test: Reject?};
 \node[decision,below=0.5cm of m2] (test2) {Test: Reject?};
 \node[decision,below=0.5cm of m3] (test3) {Test: Reject?};
 \path[line] (ols)    -> (test1);
 \path[line] (test1)  -> (gmm) ;
 \path[line] (gmm)    -> (test2);
 \path[line] (test2)  -> (sgmm) ;
 \path[line] (sgmm)   -> (test3);
 \path[line]  (test3) -> (sgmmma);
 
\end{tikzpicture}
}

\caption{Modelling strategy of a hypothetical analyst\label{fig:strategy}}
\end{figure}
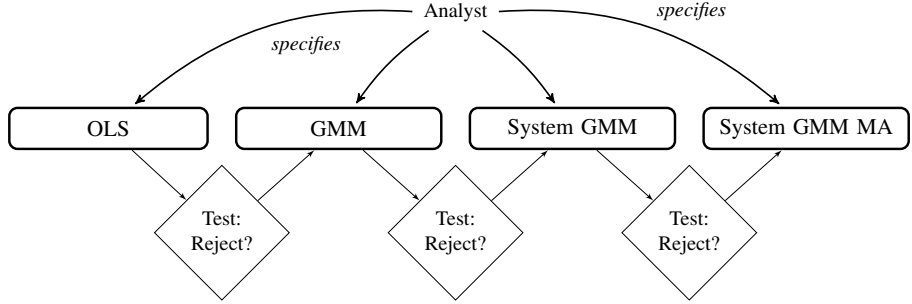

To do so, our analyst would start by using a base model (typically OLS), and then let economic and statistical theory suggest more appropriate models. In this case, we will estimate variants of a dynamic panel data model for the evolution of employment. For each model, tests can be specified to check whether the model is an appropriate fit under a certain hypothesis.\footnote{We do not describe these models in more detail here, referring the reader to the literature instead, in particular \textcite{RePEc:eee:econom:v:68:y:1995:i:1:p:29-51} and \textcite{RePEc:eee:econom:v:87:y:1998:i:1:p:115-143}.} The outcome of this exercise, illustrated by Figure~\ref{fig:strategy}, allows us to assess whether the synthetic data capture variability in economic growth due to industry, firm age and payroll --- the key variables in the data --- and whether the analyst might reasonable choose the same, or a closely related modelling strategy.

The base model is an OLS specification:
\begin{eqnarray}	
\label{eq:OLS}
Emp_{et} & = & \beta_0 + \theta Emp_{e,t-1} + \eta Pay_{et} + Age_{et}^{T}\beta + \gamma_i + \lambda_t + \epsilon_{et}
\end{eqnarray}
where $Emp_{et}$ is log employment of entity $e$ in year $t$, $Emp_{e,t-1}$ is its one year lag, $Pay_{et}$ is the logarithm of payroll of entity $e$ in year $t$, $Age_{et}$ is a vector of dummy variables for age of entity $e$ in year $t$, $\lambda_t$ is a year  effect, $\gamma_i$ is a time-invariant industry-specific effect for each industry $i$, and $\epsilon_{et}$ is the disturbance term of entity $e$ in year $t$. 
As $Emp_{e,t-1}$ is correlated with $\gamma_{i}$ because $Emp_{e,t-1}$ is itself determined by time-invariant $\gamma_{i}$, OLS estimators are biased and inconsistent. To obtain consistent estimates of the parameters in the model, \textcite{RePEc:oup:restud:v:58:y:1991:i:2:p:277-297.} suggest using  generalized method of moments (GMM) estimation methods, as well as associated tests to assess the  validity of the model.  
We also estimate the model using system GMM methods proposed by \textcite{RePEc:eee:econom:v:68:y:1995:i:1:p:29-51} and \textcite{RePEc:eee:econom:v:87:y:1998:i:1:p:115-143} (System GMM), as well as a variant of equation (\ref{eq:OLS}) that includes a first-order moving average in the error term $\epsilon_{et}$ (System GMM MA):
\begin{eqnarray}	
Emp_{et}&=&\beta_{0} +\theta Emp_{e,t-1}+\eta Pay_{et} + Age_{et}^{T}\beta + \lambda_t +\alpha_e   + \epsilon_{et} + \epsilon_{e,t-1}
\end{eqnarray}
where $\alpha_e$ is a time-invariant entity effect, which includes any time-invariant industry effects.

The Sargan test \parencite{hansen_large_1982,RePEc:oup:restud:v:58:y:1991:i:2:p:277-297.,blundell_estimation_2001} is used to assess the validity of the over-identifying restrictions. We also compute the z-score for the $m2$ test for zero autocorrelation in the  first-differenced errors of order two \parencite{RePEc:oup:restud:v:58:y:1991:i:2:p:277-297.}. 

An interesting derived effect is to consider the long-run effect of (log) payroll on (log) employment, or the elasticity of employment with respect to payroll. This can be estimated as
$$
\eta^\star = \frac{\hat{\eta}}{1-\hat{\theta}}.
$$

It is important that this model is close, but not identical to the model used to \textit{synthesize} the data. In \SynLBD, $Emp_{et}$ is synthesized as $f(Emp_{e,t-1},X_{et})$ (where $X_{et}$ does not contain $Pay_{et}$), and $Pay_{et} = f(Pay_{e,t-1},Emp_{et},X_{et})$ \citep[pg. 366]{KinneyEtAl2011}. Thus, the model we chose is purposefully not (completely) congenial with the synthesis model, but the synthesis process of the \SynLBD{} should preserve sufficient serial correlation in the data to be able to estimate these models.

\begin{figure} [H]
\centering
\includegraphics[width=\linewidth]{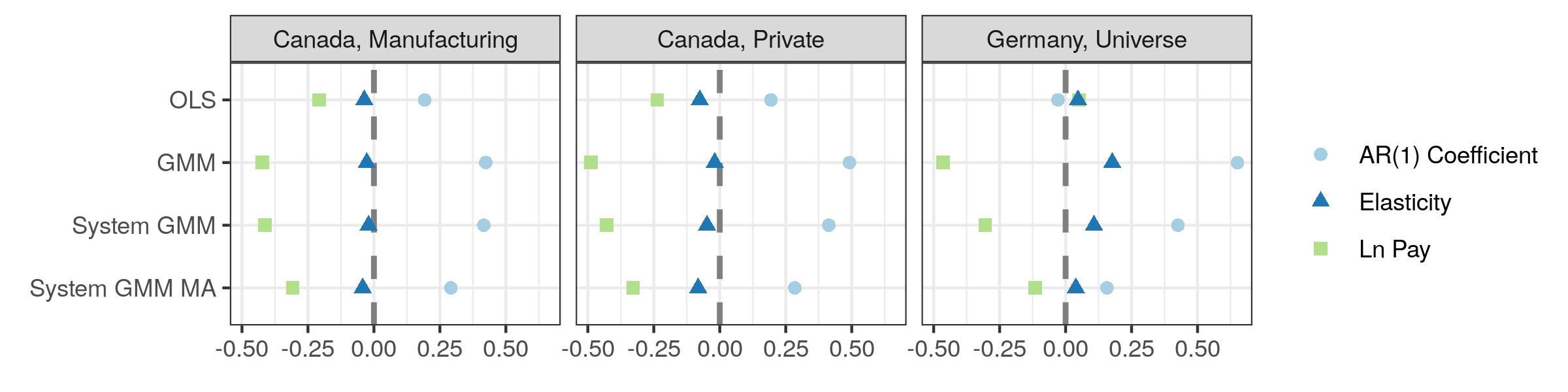}
\caption{Bias in estimates of coefficients on  pay and lagged employment\label{fig:estimates3}} 
\begin{minipage}{0.70\linewidth}
{\footnotesize \textit{Note}: For details on the estimated coefficients, see the  Appendix. \par}
\end{minipage}
\end{figure}

We estimate each model and test statistics separately on confidential and synthetic data for the private sector (and for Canada, for the manufacturing sector). Detailed estimation results are reported in the  Appendix. Here we  focus on the two regression coefficients of major interest: $\theta$ and $\eta$, the coefficients for lagged employment and payroll, as well as the elasticity $\eta^\star$. Figure~\ref{fig:estimates3} plots the bias in the synthetic coefficients, i.e., $\theta_{synth}-\theta_{conf}$ and $\eta_{synth}-\eta_{conf}$, for all four models. While the detailed results in the  Appendix confirm that all regression coefficients still have the same sign, all estimates plotted in Figure~\ref{fig:estimates3} show substantial bias in all models in all datasets (the OLS model for the German data being the only exception). Still, the computed elasticity $\eta^\star$ has very little bias in most models.

However, we observe a striking pattern: The biases of the two regression coefficients are always symmetric, i.e. the sum of the biases of $\theta_{synth}$ and $\eta_{synth}$ is close to zero in all models (and mostly cancel out in the computation of $\eta^\star$). 
This may simply be a feature of the modeling strategy pointed out earlier, which generates serial correlation with a slightly different structure. Another possible explanation could be that the model is poorly identified because of multicollinearity generating a ridge for the estimated coefficients. The estimated coefficients would be highly unstable in this case even in the original data and thus it would not be surprising to find substantial differences between the coefficients from the original data and the coefficients from the synthetic data. Better understanding this phenomenon will be an interesting area of future research.

\begin{table}[!htbp] \centering 
  \caption{m2 and Sargan tests by country} 
  \label{tab:m2sargan} 
  \resizebox{\columnwidth}{!}{

\begin{tabular}{@{\extracolsep{5pt}} llcccc} 
\\[-1.8ex]
\toprule
      &      &\multicolumn{2}{c}{Canada}&\multicolumn{2}{c}{Germany}\\
Model & Test & Confidential & Synthetic & Confidential & Synthetic \\ 
\toprule
GMM & m2 & -14.5 & -27.54 & -2.51 & -4.13 \\ 
    & Sargan test & 69000 & 15000 & 3600 & 2000 \\ 
    \midrule
System GMM & m2 & -11.43 & -41.6 & 19.49 & -8.83 \\ 
           & Sargan test & 77000 & 18000 & 4500 & 2800 \\ 
           \midrule
System GMM MA & m2 & 8.2 & -40.03 & 19.03 & -11.69 \\ 
              & Sargan test & 28000 & 17000 & 3100 & 2500 \\ 
\bottomrule
\end{tabular} 
}
\begin{tablenotes}\footnotesize \item 
\textit{Note}: The Sargan test \parencite{blundell_estimation_2001,RePEc:oup:restud:v:58:y:1991:i:2:p:277-297.} is used to assess the validity of the over-identifying restrictions. The z-score for the $m2$ tests for zero autocorrelation in the  first-differenced errors of order two \parencite{RePEc:oup:restud:v:58:y:1991:i:2:p:277-297.}. See text for additional information.
\end{tablenotes}

\end{table}

While the bias in coefficients is quite consistent across countries and models, specification tests such as  the $m2$ test for autocorrelation and the Sargan test paint a slightly less consistent picture. Table~\ref{tab:m2sargan} shows the two tests for each of the models estimated by country, synthetic status, and model. The Sargan test rejects the null in both countries and for all models, consistently for confidential and synthetic data. But the $m2$ test is of opposite signs for half of the comparisons.

\subsection{pMSE}

To compute the $pMSE$, we estimate Equation (\ref{pMSE}) using logit models. The estimated $pMSE$ is 0.0121 for the Canadian data (0.0041 for the manufacturing sector) and 0.0013 for the German data (see Table~\ref{tab:pmse}). While these numbers may seem small, the $pMSE$ ratio and the standardized $pMSE$ are large, indicating that the null hypothesis that the synthetic data and the original data stem from the same data generating process should be rejected. The expected $pMSE$ is quite sensitive to sample size $N$. Even small differences between the original and synthetic data will lead to large values for this test statistic. In both countries, the confidential data files are quite large (about 2 million cases for Germany and the manufacturing sector in Canada and about 34.5 million cases for the full Canadian data sets). In practice, therefore, it is quite likely to reject the null of equivalence  given this test's very high power.

\begin{table}[!htbp] \centering 
  \caption{pMSE by sector and country} 
  \label{tab:pmse} 
\begin{tabular}{@{\extracolsep{5pt}} llccc} 
\toprule
Country & Sector & pMSE & pMSE ratio & standardized pMSE \\ 
\midrule
Canada & Manufacturing & 0.0041 & 656.88 & 4908.17 \\ 
Canada & Private & 0.0121 & 10957.61 & 135525.77 \\ 
Germany & Universe & 0.0013 & 725.21 & 2896.85 \\ 
\bottomrule
\end{tabular} 
\end{table}

\section{Confidentiality protection}
\label{sec:confidentiality}

\begin{figure}[ht]
\includegraphics[width=\linewidth]{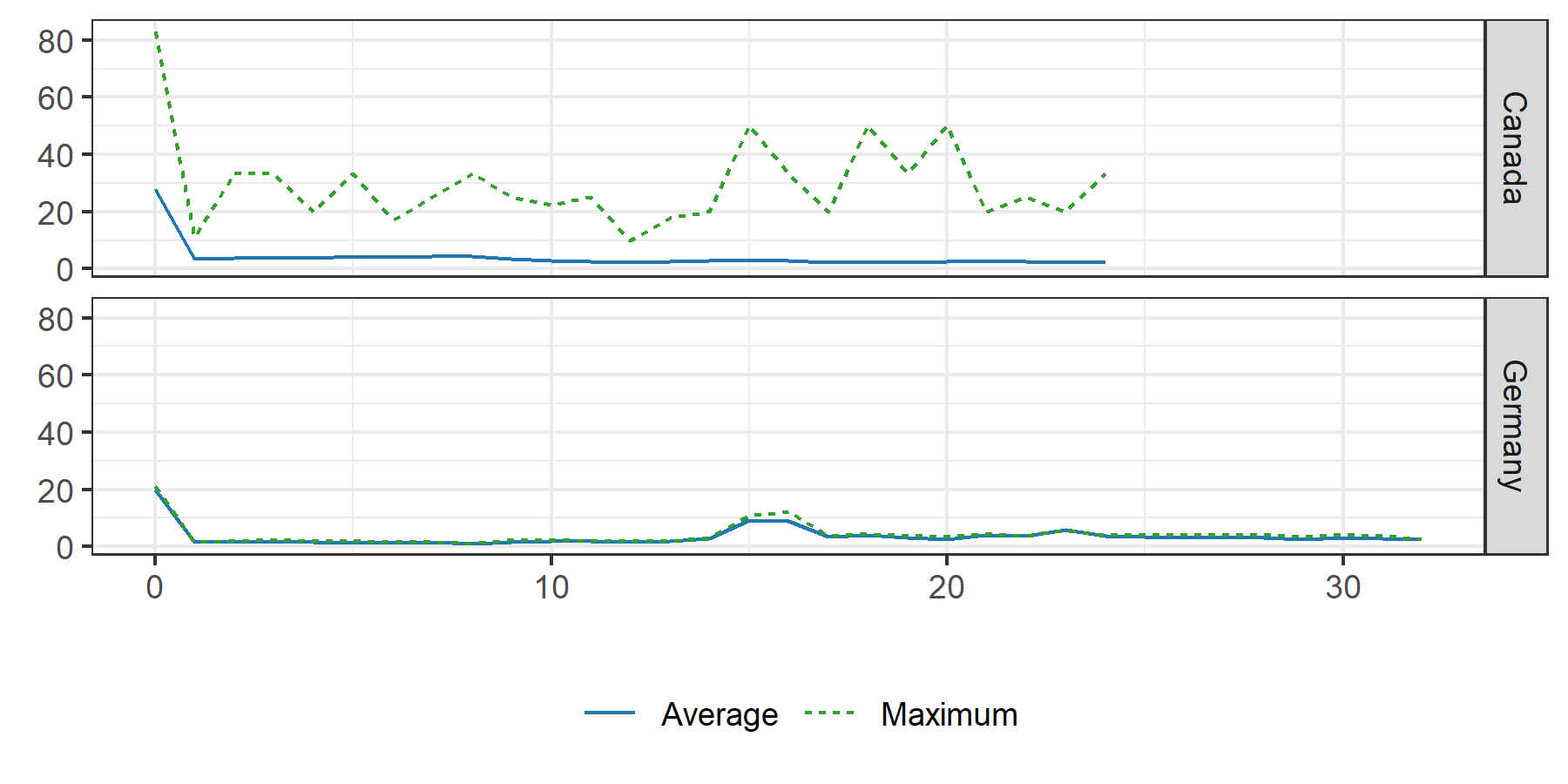}
\caption{Average and maximum likelihood that synthetic birthyear matches actual birthyear\label{fig:Conf.Both}}

\begin{center}
\begin{minipage}{0.7\linewidth}
	\footnotesize \it
Note: Plot shows fraction of entities by industry for which the synthetic birth year equals the true birth year, conditional on the synthetic birth year. Plot has been rescaled to be relative to the first year observed in the data.	
\end{minipage}
\end{center}
\end{figure}

To assess the risk of disclosure, we use a measure proposed by \textcite{KinneyEtAl2011}: For each industry, we estimate the fraction of entities for which the synthetic birth year equals the true birth year, conditional on the synthetic birth year, and interpret it as a probability. Tables 14 and 15 
in the  Appendix show the minimum, maximum, and mean of these probabilities, by year. Figure~\ref{fig:Conf.Both} shows the maximum and average values across time, for each country.\footnote{The Canadian manufacturing sector is not shown. In the German case, we only use two industries, but we show the average of the two, rather than the values for both industries, to maintain comparability with the Canadian plot.} The figure shows that these probabilities are quite low except for the first year. Entry rates in the first year are much larger than in any other year due to censoring. It is therefore quite likely that the (left-censored) entry year of the synthetic record matches that of the (left-censored) original record if the synthetic entry year is the first year observed in the data. A somewhat more muted version of this effect can be seen for Germany in the years 1991 and 1992, when the lower panel of Figure~\ref{fig:Conf.Both} shows another spike. These are the years in which data from Eastern Germany were added to the database successively, leading to new sets of (left-censored) entities. 

With the exception of the first year in the data, the average rate of concordance between synthetic and observed birth year of an establishment in the Canadian data is below 5\%, and the maximum is never above 50\%. The German data reflect results from a smaller set of industries, and while the average concordance is higher (never above 10\%), the maximum is never above 6\% other than during the noted entry spikes. This suggests that the synthetic lifespan of any given entity is highly unlikely to be matched to its confidential real lifespan. This is generally considered to be a high degree of confidentiality.

 \newpage
\section{Conclusion}
\label{sec:conclusion}

In this paper, we presented results from two projects that evaluated whether the code developed to synthesize the U.S. LBD can easily be adapted to create synthetic versions of similar data from Canada and Germany. We considered both univariate time-series comparisons as well as model-based comparisons of coefficients and model fit. In general, utility evaluations show significant differences between each country's synthetic and confidential data. Frequently-used measures such as confidence interval overlap and $pMSE$ suggest that the synthetic data are an unreliable image of the confidential data. Less formal comparisons of specification test scores suggest that the synthetic data do not reliably lead to  the same modeling decisions.

Interestingly, the utility of the German synthetic data was higher than the utility of the Canadian data in almost all dimensions evaluated. At this point we can only speculate about potential reasons. The most important difference between the two data sources is that the German data comprises only a handful of industries while almost all industries have been included in the Canadian evaluation. Given that the industries included in the German data were rather large, and synthesis models are run independently for each industry, it might have been easier to preserve the industry level statistics for the German data. We cannot exclude the possibility that  the structure of the German data aligns more closely with the LBD and thus the synthesis models tuned on the LBD data provide better results on the (adjusted) BHP than on the LEAP. We note that both the LBD and the BHP are establishment-level data sets, whereas the LEAP is an employer-level data set. 

We emphasize that adjustments to the original synthesis code were explicitly limited to ensuring that the code runs on the new input data. The validity of the synthetic data could possibly be improved by tuning the synthesis models to the particularities of the data at hand, such as the non-standard dynamics introduced into the German data by reunification.  However, the aim of this project was to illustrate that the high investments necessary for developing the synthesis code for the LBD offered additional payoffs as the re-use of the code substantially reduced the amount of work required to generate decent synthetic data products for other business data. One of the major criticisms of the synthetic data approach has been  that investments necessary to develop useful synthesizers are substantial. This project illustrated that substantial gains can be achieved when exploiting knowledge from previous projects. With the advent of tailor-made software such as the \textit{synthpop} package in R \citep{JSSv074i11}, the investments for generating useful synthetic data might be further reduced in the future.

However, even without  fine-tuning or customization of models, the current synthetic data have, in fact, proven useful. De facto, many deployments of synthetic data, including the Synthetic LBD in the US, have been used for model preparation by researchers in a public or lower-security environment, with subsequent remote submission of prepared code for validation against the confidential data. When viewed through the lens of such a validation system, the synthetic data prepared here would seem to have reasonable utility. While time series dynamics are not the same, they are broadly similar. Models converged in similar fashions, and while coefficients were strictly different, they were broadly similar and plausible. Specification tests did not lead to the same conclusions, but they also did not collapse or yield meaningless conclusions. Thus, we believe that the synthetic data, despite being different, have the potential to be a useful tool for analysts to prepare models without direct access to the confidential data. \textcite{VilhuberAbowd2016-SOLE,Vilhuber2019-SGP} come to a similar conclusion when evaluating usage of the synthetic data sets available through the Synthetic Data Server \citep{AbowdVilhuber2010}, including the Synthetic LBD. A more thorough evaluation would need to explicitly measure the investment in synthetic data generation, the cost of setting up a validation structure, and the number of studies enabled through such a setup. We note that such an evaluation is non-trivial: the counter-factual in many circumstances is that no access is allowed to sensitive business microdata, or that access occurs through a secure research data system that is also costly to maintain. This study has contributed to such a future evaluation by showing that plausible results can be achieved with relatively low up-front investments.
 
The use of synthetic data sets to broaden access to confidential microdata is likely to increase in the near future, with increasing concerns by statistical agencies regarding the disclosure risks of releasing microdata. The resulting reduction in access to scientific microdata is overwhelmingly seen as problematic. Broadly ``plausible'' if not analytically valid synthetic data sets such as those described in this paper, combined with scalable remote submission systems that integrate modern disclosure avoidance mechanisms, may be a feasible mitigation strategy.

\section*{Acknowledgements}

The opinions expressed here are those of the authors, and do not reflect the opinions of any of the statistical agencies involved. All results were reviewed for disclosure risks by their respective custodians, and released to the authors.  Alam thanks Claudiu Motoc and Danny Leung for help with the Canadian data.  Vilhuber acknowledges funding through NSF Grants SES-1131848 and SES-1042181, and a grant from Alfred P. Sloan Grant (G-2015-13903). Alam and Dostie acknowledge funding through SSHRC Partnership Grant ``Productivity, Firms and Incomes''. The creation of the Synthetic LBD  was funded by NSF Grant SES-0427889.

\printbibliography

\include{empty}

\include{online-appendix-final}
  \end{document}

%% file: empty.tex

%% file: online-appendix-final.tex
\newpage

\FloatBarrier

\appendix

\begin{center}
{\LARGE{Appendix}}

{``Applying Data Synthesis for Longitudinal Business Data across Three Countries''}

\textit{M. Jahangir Alam, Benoit Dostie, J\"org Drechsler, Lars Vilhuber}

\end{center}

\section{Figures for the Manufacturing Sector in Canada}
\label{sec:appendix_figures}

\begin{figure}[H]
\centering
\begin{subfigure}[h]{0.48\linewidth}
\label{tab:Can:GrossEmploymentPrivate}
\includegraphics[width=\linewidth]{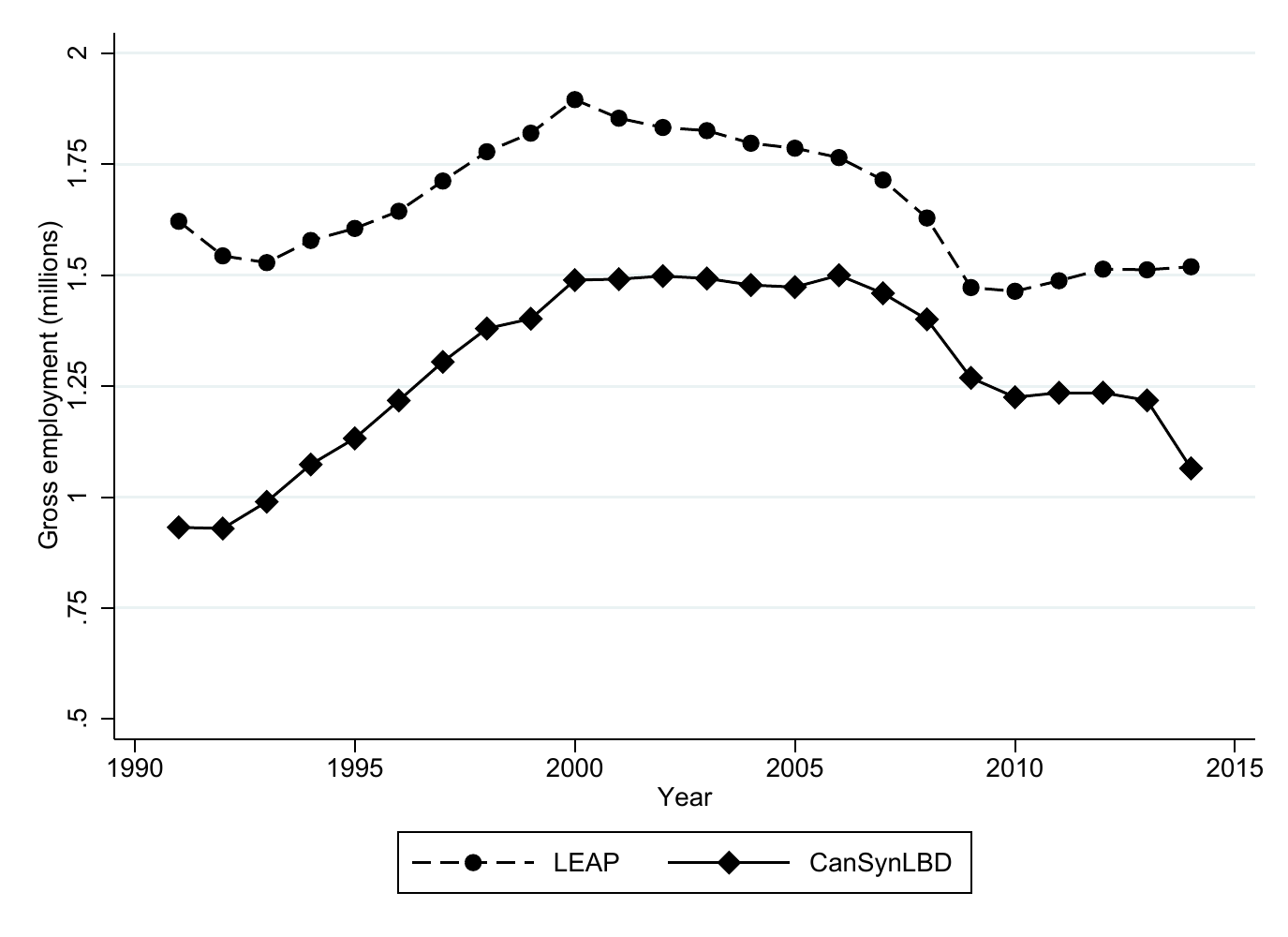} 
\caption{Gross employment level by year} \end{subfigure}
\hfill
\begin{subfigure}[h]{0.48\linewidth}
\includegraphics[width=\linewidth]{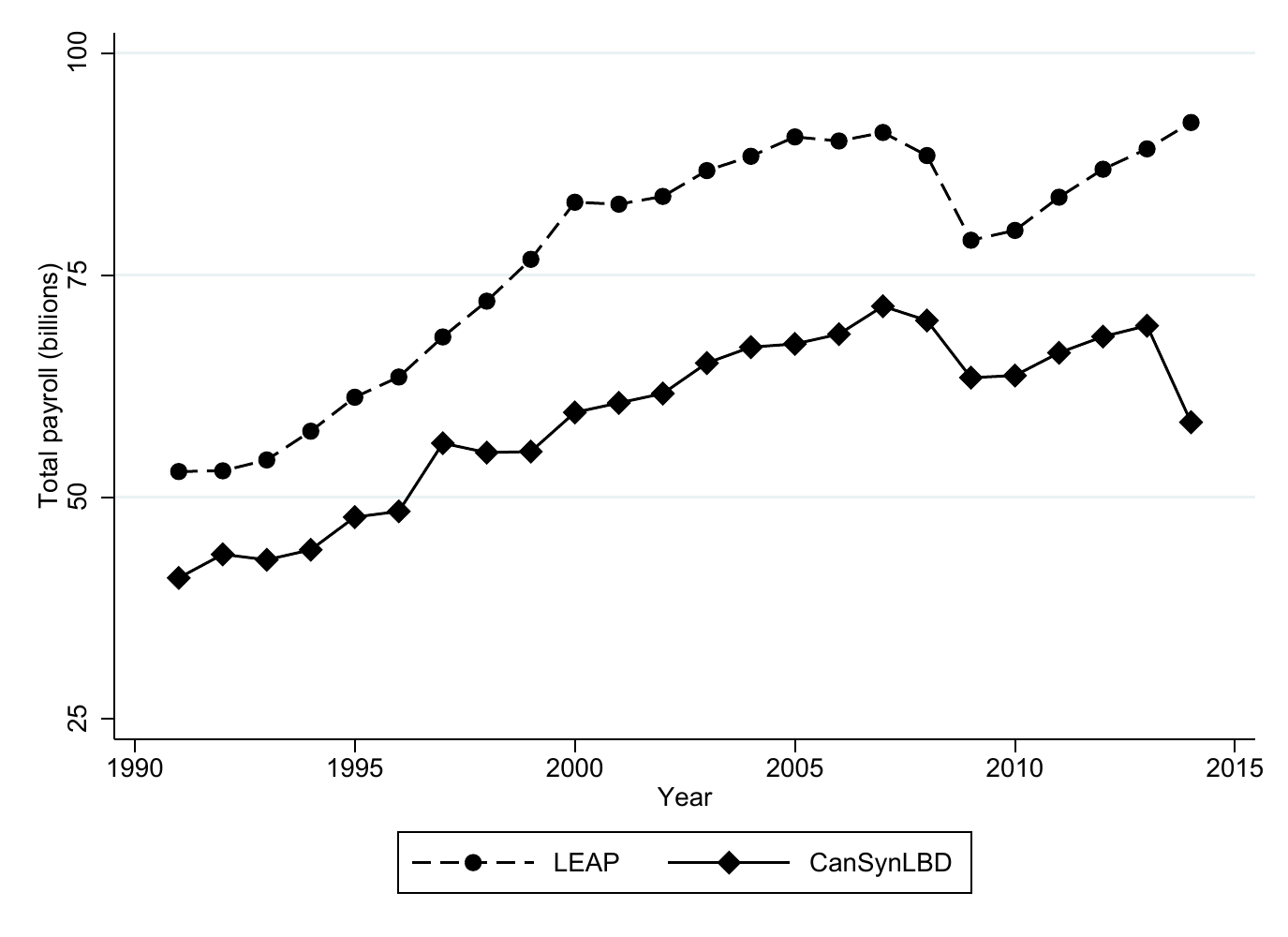}
\caption{Total payroll}
\end{subfigure}\caption{Entity characteristics for the manufacturing sector in Canada by year.}\label{fig:entity_chracteristics_manufac}
\end{figure}

\begin{figure}[H]
\begin{subfigure}[h]{0.48\linewidth}
\includegraphics[width=\linewidth]{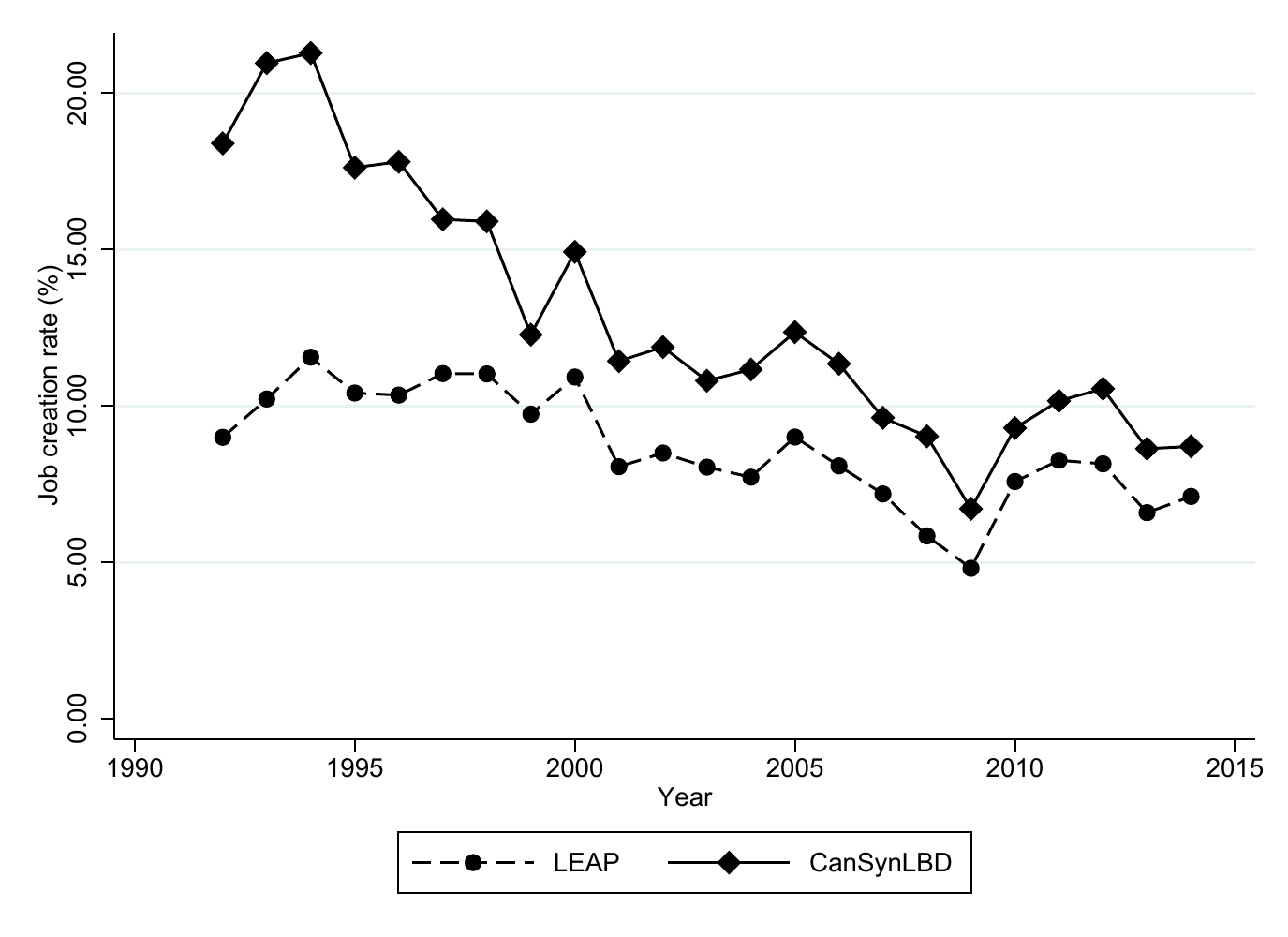}
\caption{Job creation rates}
\end{subfigure}
\hfill
\begin{subfigure}[h]{0.48\linewidth}
\includegraphics[width=\linewidth]{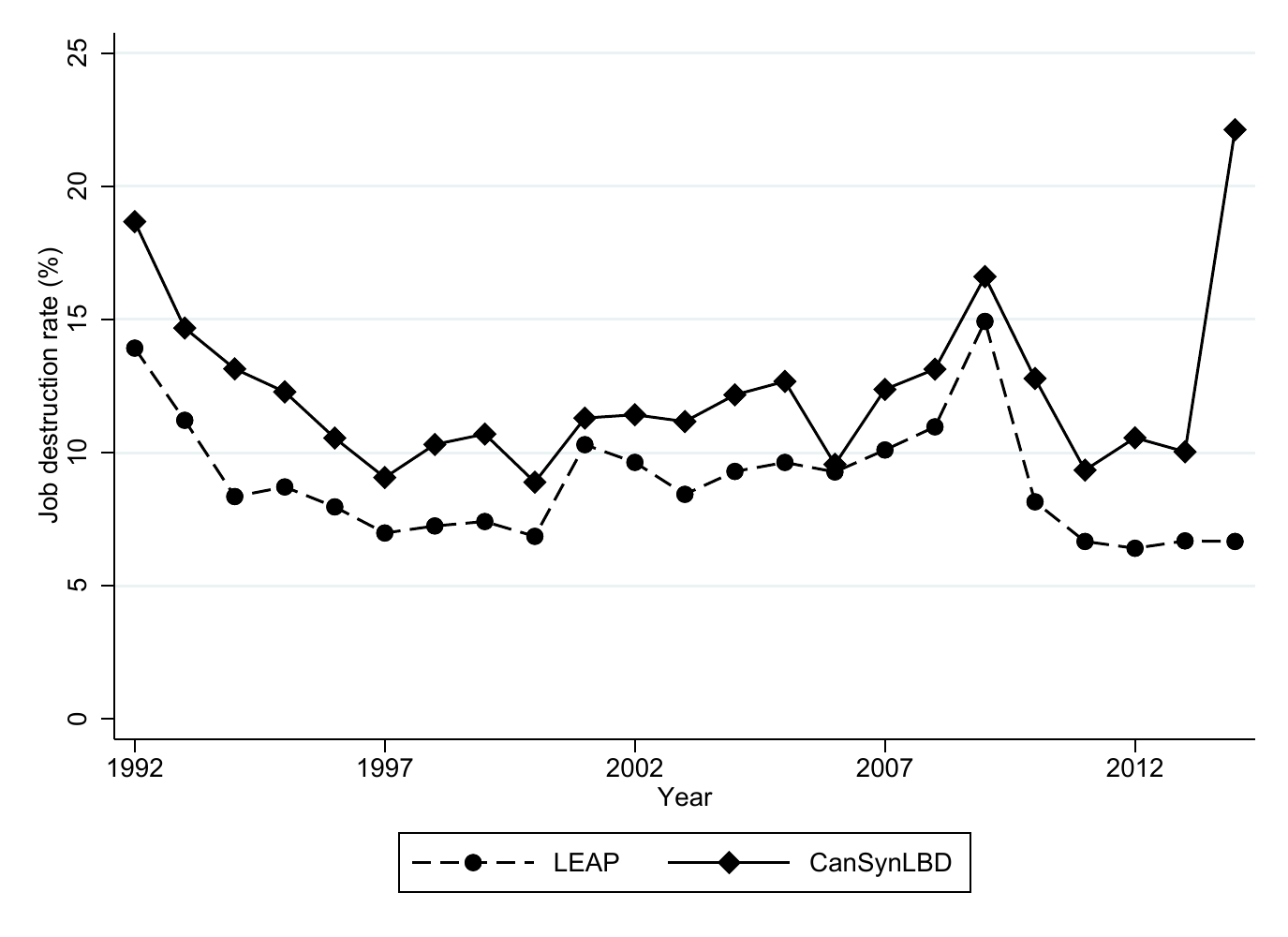}
\caption{Job destruction rates}
\end{subfigure}\caption{Dynamics of job flows for the manufacturing sector in Canada by year.}\label{fig:job_flows_manufac}
\end{figure}

\begin{figure}[H]
\centering
\begin{subfigure}[h]{0.5\linewidth}
\includegraphics[trim=0 10 0 0,clip, width=\linewidth]{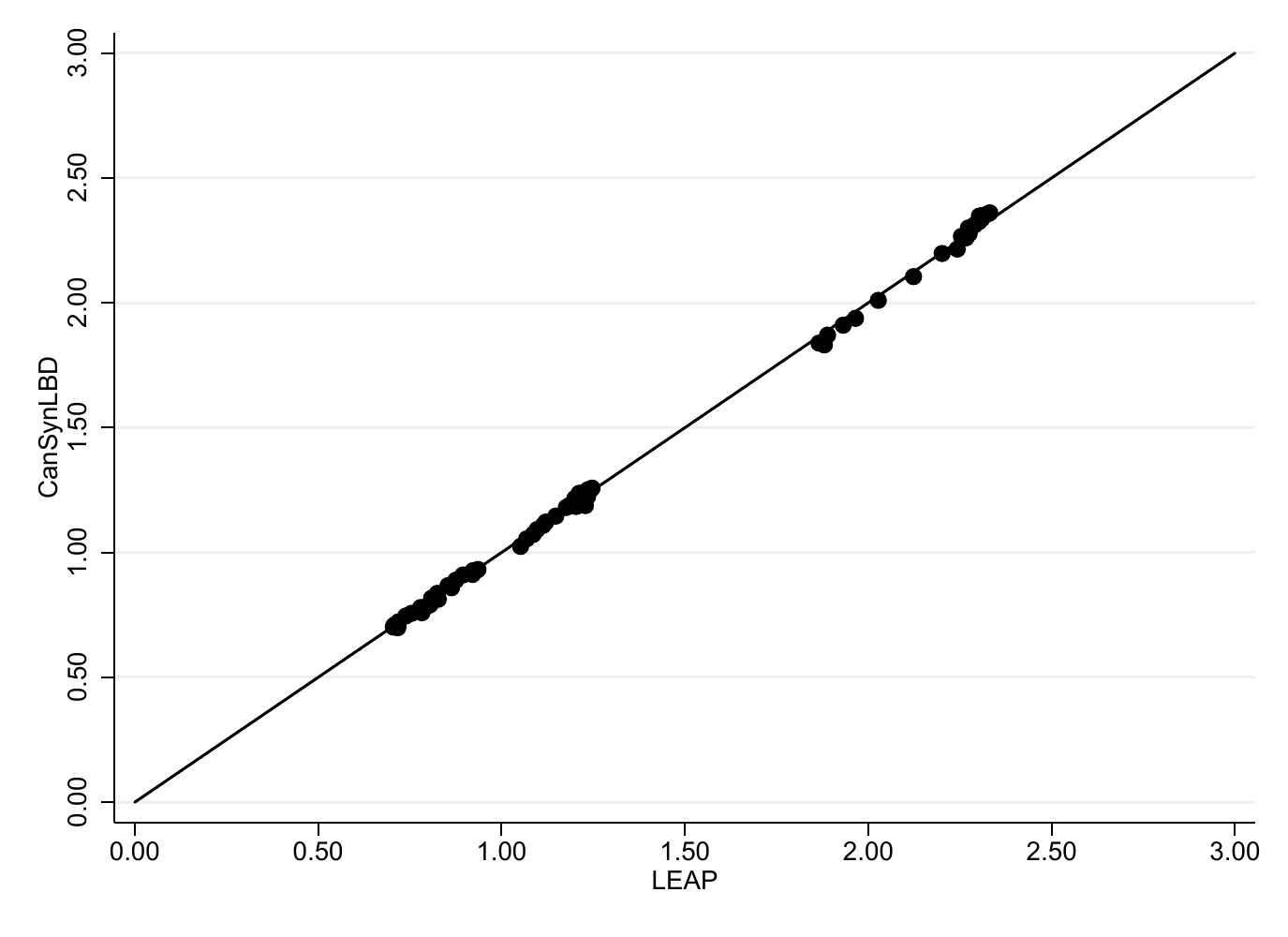}
\end{subfigure}\\
\begin{subfigure}[h]{0.5\linewidth}
\includegraphics[trim=0 10 0 -20,clip,width=\linewidth]{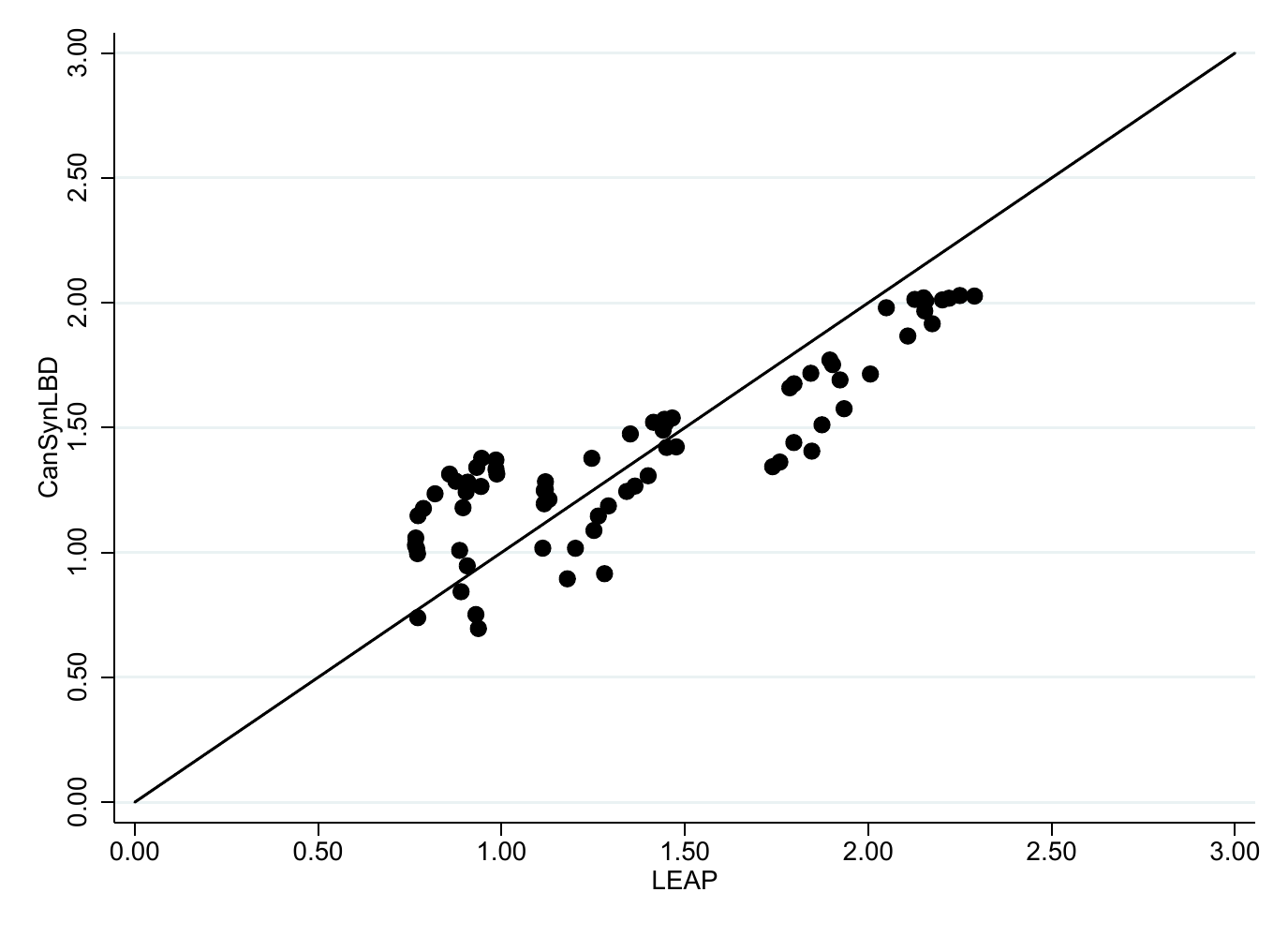}
\end{subfigure}\\
\begin{subfigure}[h]{0.5\linewidth}
\includegraphics[trim=0 0 0 -20,clip,width=\linewidth]{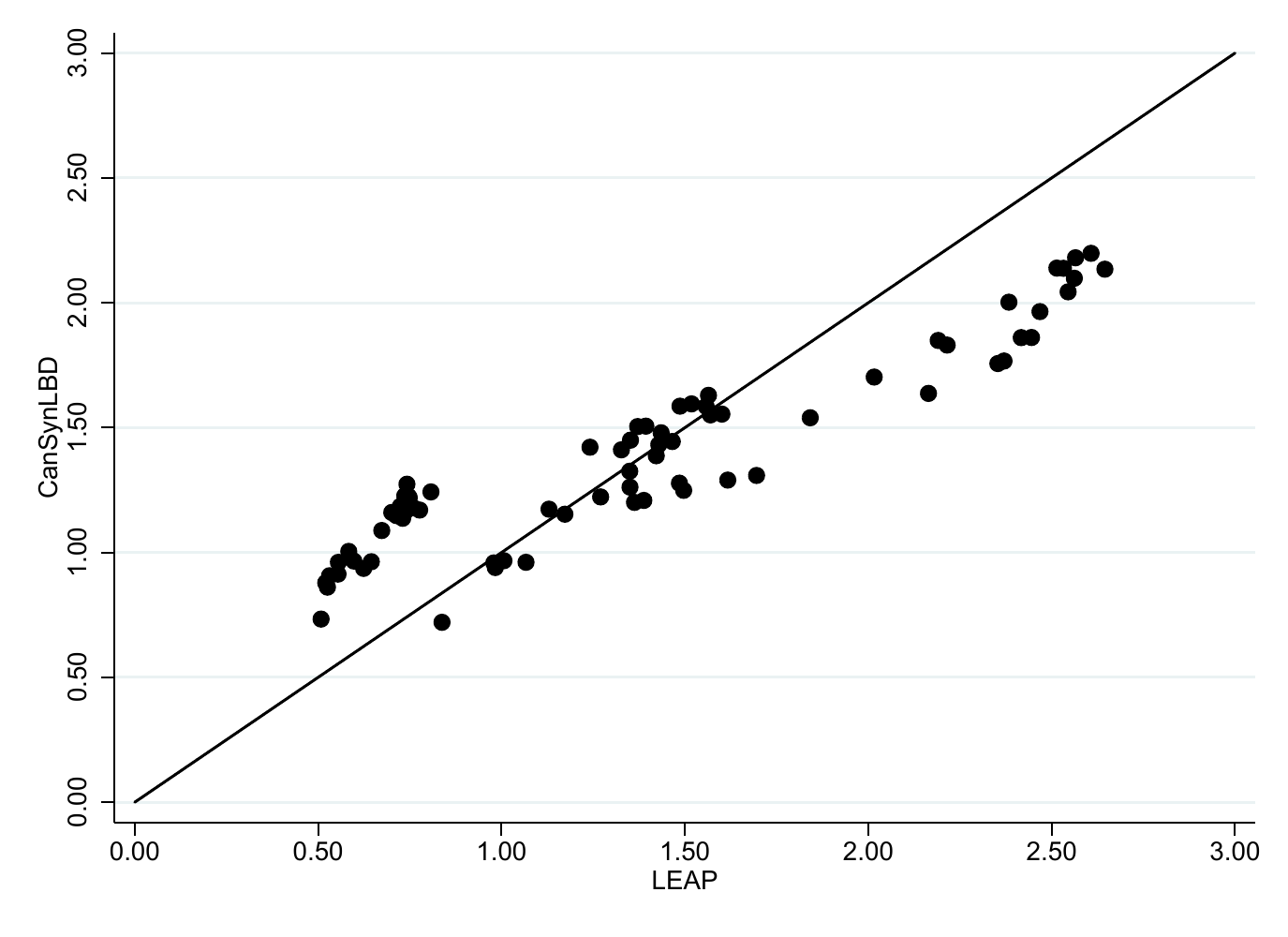}
\end{subfigure}
\caption{Share of entities (upper panel), share of employment (middle panel), and share of payroll (lower panel) by year and industry for the Canadian manufacturing sector.}\label{fig:FirmShare_manufac}
\end{figure}

\FloatBarrier
\clearpage

\section{Appendix Tables}
\label{sec:appendix_tables}

\subsection{pMSE}
\label{sec:pmse_tables}

\begin{table}[!htbp] \centering 
\setlength{\tabcolsep}{11pt}
\begin{threeparttable}
  \caption{Detailed results for pMSE estimation by sector and country} 
  \label{tab:pmse:details} 
\begin{tabular}{@{\extracolsep{5pt}} l|cc|c} 
\toprule
\textbf{Independent Variables} & \multicolumn{2}{c|}{\textbf{Canada}} & \textbf{Germany}\\
\multicolumn{1}{r|}{\it Sector:}&Manufacturing & Private & All \\ 
\midrule
Ln ALU & 0.158 & 0.7138 & -0.2895 \\ 
 & (0.0039) & (0.001) & (0.0033)\\
Ln Pay & 0.0039 & -0.4426 & 0.2584 \\ 
 & (0.0037) & (0.001) & (0.0028)\\
Age 3-4 & 0.0392 & 0.0972 & -0.0987 \\ 
 & (0.0078) & (0.0017) & (0.007)\\ 
Age 5-7 & -0.0382 & 0.0477 & -0.0973 \\ 
 & (0.0073) & (0.0016) & (0.0066)\\ 
Age 8-12 & -0.1258 & -0.0263 & -0.1172 \\ 
 & (0.0071) & (0.0015) & (0.0063)\\ 
Age 13 or more & -0.219 & -0.1024 & -0.1487 \\ 
 & (0.0074) & (0.0016) & (0.0059)\\ 
 \midrule
N & 2243011 & 34638723 & 2121956 \\ 
pseudo R-sq & 0.0112 & 0.0318 & 0.0038 \\ 
\midrule
pMSE & 0.0041 & 0.0121 & 0.0013 \\ 
\bottomrule
\end{tabular} 
\begin{tablenotes}
\small
\item \textit{Note}: See Equation~\ref{pMSE} for estimation method. An observation is a entity-year in the combined database of each country-sector combination. All specifications include  time and industry fixed effects. Standard errors are in parentheses. 
\end{tablenotes}
\end{threeparttable}
\end{table} 

\clearpage
\FloatBarrier
\clearpage
 
\subsection{Regression analysis tables}
\label{sec:regression_tables}

\begin{table}[H]
  \centering
\begin{threeparttable}
 \caption{Regression coefficients (OLS) for LEAP} \label{tab:OLS_can} \medskip
\renewcommand{\arraystretch}{1}
\begin{tabular}{l|c c| c c}
\toprule
\textbf{Independent Variables}&\multicolumn{2}{c|}{\textbf{LEAP}} &  \multicolumn{2}{c}{\textbf{CanSynLBD}}\\
\midrule
&\multicolumn{1}{c}{Private}&\multicolumn{1}{c}{Manufacturing}&\multicolumn{1}{c}{Private}&\multicolumn{1}{c}{Manufacturing}\\
\hline
AR(1) Coefficient&   0.2031\sym{***}&   0.2481\sym{***}&   0.3970\sym{***}&   0.4405\sym{***}\\
          & (0.0001)         & (0.0005)         & (0.0002)         & (0.0007)         \\
[1em]
Ln Pay    &   0.7847\sym{***}&   0.7300\sym{***}&   0.5481\sym{***}&   0.5228\sym{***}\\
          & (0.0001)         & (0.0005)         & (0.0002)         & (0.0006)         \\
[1em]
Age 3-4   &  -0.1202\sym{***}&  -0.1717\sym{***}&  -0.1223\sym{***}&  -0.2340\sym{***}\\
          & (0.0003)         & (0.0014)         & (0.0004)         & (0.0016)         \\
[1em]
Age 5-7   &  -0.1260\sym{***}&  -0.1891\sym{***}&  -0.1235\sym{***}&  -0.2507\sym{***}\\
          & (0.0003)         & (0.0014)         & (0.0004)         & (0.0016)         \\
[1em]
Age 8-12  &  -0.1268\sym{***}&  -0.1973\sym{***}&  -0.1169\sym{***}&  -0.2551\sym{***}\\
          & (0.0003)         & (0.0013)         & (0.0004)         & (0.0016)         \\
[1em]
Age 13 or more&  -0.1246\sym{***}&  -0.1992\sym{***}&  -0.1101\sym{***}&  -0.2577\sym{***}\\
          & (0.0003)         & (0.0014)         & (0.0004)         & (0.0017)         \\
\hline
\(N\)     & 15708195         &  1015293         & 13573225         &   959764         \\
\(R^{2}\) &   0.9696         &   0.9743         &   0.9444         &   0.9523         \\
 \bottomrule
\end{tabular} 
\begin{tablenotes}
\small
\item Note: In all specifications, we include both year and industry fixed effects. Standard errors are in parentheses.  ***, **, and * indicate statistically significant coefficients at 1\%, 5\%, and 10\% percent levels, respectively.
 \end{tablenotes}
 \end{threeparttable}
\end{table}

\begin{table}[H]
  \centering
\caption{Regression coefficients (OLS) for GLBD} \label{tab:OLS_ger} \medskip
\renewcommand{\arraystretch}{1}
\setlength{\tabcolsep}{14pt}
\begin{tabular}{l|c |c}
\toprule
\textbf{Independent Variables}&\textbf{GLBD} &  \textbf{GSynLBD}\\
\midrule

AR(1) Coefficient&   0.4430\sym{***}&   0.4143\sym{***}\\
          & (0.0007)         & (0.0008)         \\
[1em]
Ln Pay    &   0.4629\sym{***}&   0.5143\sym{***}\\
          & (0.0006)         & (0.0007)         \\
[1em]
Age 3-4   &  -0.0695\sym{***}&  -0.0642\sym{***}\\
          & (0.0017)         & (0.0016)         \\
[1em]
Age 5-7   &  -0.1066\sym{***}&  -0.0891\sym{***}\\
          & (0.0017)         & (0.0016)         \\
[1em]
Age 8-12  &  -0.1324\sym{***}&  -0.1109\sym{***}\\
          & (0.0017)         & (0.0016)         \\
[1em]
Age 13 or more&  -0.1880\sym{***}&  -0.1600\sym{***}\\
          & (0.0016)         & (0.0015)         \\
\midrule
\(N\)     &   848871         &   966084         \\
\(R^{2}\) &   0.9167         &   0.8968         \\
    \bottomrule
  \end{tabular} 
\begin{tablenotes}
\small
\item Note: In all specifications, we include both year and industry fixed effects. Standard errors are in parentheses.  ***, **, and * indicate statistically significant coefficients at 1\%, 5\%, and 10\% percent levels, respectively.
 \end{tablenotes}
\end{table}

\begin{table}[H]
  \centering
\begin{threeparttable}
 \caption{Regression coefficients (Dynamic) for LEAP} \label{tab:Dynamic - GMM_can} \medskip
\renewcommand{\arraystretch}{1}
\begin{tabular}{l|c c| c c}
\toprule
\textbf{Independent Variables}&\multicolumn{2}{c|}{\textbf{LEAP}} &  \multicolumn{2}{c}{\textbf{CanSynLBD}}\\
\midrule
&\multicolumn{1}{c}{Private}&\multicolumn{1}{c}{Manufacturing}&\multicolumn{1}{c}{Private}&\multicolumn{1}{c}{Manufacturing}\\
\hline
AR(1) Coefficient&   0.0805\sym{***}&   0.1189\sym{***}&   0.5722\sym{***}&   0.5425\sym{***}\\
          & (0.0003)         & (0.0018)         & (0.0024)         & (0.0084)         \\
[1em]
Ln Pay    &   0.8991\sym{***}&   0.8523\sym{***}&   0.4101\sym{***}&   0.4302\sym{***}\\
          & (0.0002)         & (0.0015)         & (0.0018)         & (0.0067)         \\
[1em]
Age 3-4   &  -0.0450\sym{***}&  -0.0797\sym{***}&  -0.2075\sym{***}&  -0.2972\sym{***}\\
          & (0.0002)         & (0.0014)         & (0.0010)         & (0.0051)         \\
[1em]
Age 5-7   &  -0.0438\sym{***}&  -0.0860\sym{***}&  -0.2129\sym{***}&  -0.3162\sym{***}\\
          & (0.0002)         & (0.0015)         & (0.0011)         & (0.0059)         \\
[1em]
Age 8-12  &  -0.0418\sym{***}&  -0.0923\sym{***}&  -0.2187\sym{***}&  -0.3294\sym{***}\\
          & (0.0003)         & (0.0017)         & (0.0013)         & (0.0070)         \\
[1em]
Age 13 or more&  -0.0379\sym{***}&  -0.0898\sym{***}&  -0.2318\sym{***}&  -0.3414\sym{***}\\
          & (0.0003)         & (0.0019)         & (0.0015)         & (0.0080)         \\
\hline
\(N\)     & 15708195         &  1015293         & 13573225         &   959764         \\
m2        & -14.5000         &  -2.2200         & -27.5400         &  -9.4400         \\
Sargan test&  6.9e+04         &  4.6e+03         &  1.5e+04         &  1.5e+03         \\
df of Sargan Test& 252.0000         & 252.0000         & 252.0000         & 252.0000         \\
P value of Sargan test&   0.0000         &   0.0000         &   0.0000         &   0.0000         \\
    \bottomrule
  \end{tabular} 
\begin{tablenotes}
\small
\item Note: In this table, $m2$ is the Arellano-Bond test for zero autocorrelation in first-differenced errors for order two. Standard errors are in parentheses. ***, **, and * indicate statistically significant coefficients at 1\%, 5\%, and 10\% percent levels, respectively.
 \end{tablenotes}
 \end{threeparttable}
\end{table}

\begin{table}[H]
  \centering
\caption{Regression coefficients (Dynamic) for GLBD} \label{tab:Dynamic - GMM_ger} \medskip
\renewcommand{\arraystretch}{1}
\setlength{\tabcolsep}{13pt}
\begin{tabular}{l|c |c}
\toprule
\textbf{Independent Variables}&\textbf{GLBD} &\textbf{GSynLBD}\\
\midrule
AR(1) Coefficient&   0.0489\sym{***}&   0.6999\sym{***}\\
          & (0.0051)         & (0.0057)         \\
[1em]
Ln Pay    &   0.7559\sym{***}&   0.2916\sym{***}\\
          & (0.0035)         & (0.0042)         \\
[1em]
Age 3-4   &  -0.0070\sym{***}&  -0.1026\sym{***}\\
          & (0.0012)         & (0.0015)         \\
[1em]
Age 5-7   &  -0.0233\sym{***}&  -0.1386\sym{***}\\
          & (0.0014)         & (0.0017)         \\
[1em]
Age 8-12  &  -0.0473\sym{***}&  -0.1694\sym{***}\\
          & (0.0015)         & (0.0018)         \\
[1em]
Age 13 or more&  -0.1084\sym{***}&  -0.2183\sym{***}\\
          & (0.0015)         & (0.0018)         \\
\hline
\(N\)     &   848871         &   966084         \\
m2        &  -2.5100         &  -4.1300         \\
Sargan test&  3.6e+03         &  2.0e+03         \\
df of Sargan Test& 495.0000         & 495.0000         \\
P value of Sargan test&   0.0000         &   0.0000         \\
    \bottomrule
  \end{tabular} 
\begin{tablenotes}
\small
\item Note: In this table, $m2$ is the Arellano-Bond test for zero autocorrelation in first-differenced errors for order two. Standard errors are in parentheses. ***, **, and * indicate statistically significant coefficients at 1\%, 5\%, and 10\% percent levels, respectively.
 \end{tablenotes}
\end{table}

\begin{table}[H]
  \centering
\begin{threeparttable}
 \caption{Regression coefficients (Dynamic - system GMM) for LEAP} \label{tab:Dynamic - system GMM_can} \medskip
\renewcommand{\arraystretch}{1}
\begin{tabular}{l|c c| c c}
\toprule
\textbf{Independent Variables}&\multicolumn{2}{c|}{\textbf{LEAP}} &  \multicolumn{2}{c}{\textbf{CanSynLBD}}\\
\midrule
&\multicolumn{1}{c}{Private}&\multicolumn{1}{c}{Manufacturing}&\multicolumn{1}{c}{Private}&\multicolumn{1}{c}{Manufacturing}\\
\hline
AR(1) Coefficient&   0.0978\sym{***}&   0.1614\sym{***}&   0.5111\sym{***}&   0.5780\sym{***}\\
          & (0.0002)         & (0.0014)         & (0.0008)         & (0.0041)         \\
[1em]
Ln Pay    &   0.8854\sym{***}&   0.8161\sym{***}&   0.4562\sym{***}&   0.4022\sym{***}\\
          & (0.0002)         & (0.0012)         & (0.0006)         & (0.0033)         \\
[1em]
Age 3-4   &  -0.0555\sym{***}&  -0.1097\sym{***}&  -0.1828\sym{***}&  -0.3177\sym{***}\\
          & (0.0002)         & (0.0012)         & (0.0004)         & (0.0028)         \\
[1em]
Age 5-7   &  -0.0558\sym{***}&  -0.1201\sym{***}&  -0.1860\sym{***}&  -0.3408\sym{***}\\
          & (0.0002)         & (0.0013)         & (0.0005)         & (0.0031)         \\
[1em]
Age 8-12  &  -0.0548\sym{***}&  -0.1298\sym{***}&  -0.1875\sym{***}&  -0.3583\sym{***}\\
          & (0.0002)         & (0.0014)         & (0.0005)         & (0.0036)         \\
[1em]
Age 13 or more&  -0.0524\sym{***}&  -0.1317\sym{***}&  -0.1943\sym{***}&  -0.3747\sym{***}\\
          & (0.0002)         & (0.0016)         & (0.0006)         & (0.0041)         \\
\hline
\(N\)     & 15708195         &  1015293         & 13573225         &   959764         \\
m2        & -11.4300         &   1.3900         & -41.6000         &  -7.6700         \\
Sargan test&  7.7e+04         &  6.3e+03         &  1.8e+04         &  1.7e+03         \\
df of Sargan Test& 274.0000         & 274.0000         & 274.0000         & 274.0000         \\
P value of Sargan test&   0.0000         &   0.0000         &   0.0000         &   0.0000         \\
    \bottomrule
  \end{tabular} 
\begin{tablenotes}
\small
\item Note: An observation is an entity-year. In this table, $m2$ is the Arellano-Bond test for zero autocorrelation in first-differenced errors for order two. Standard errors are in parentheses. ***, **, and * indicate statistically significant coefficients at 1\%, 5\%, and 10\% percent levels, respectively.
 \end{tablenotes}
 \end{threeparttable}
\end{table}

\begin{table}[H]
  \centering
\caption{Regression coefficients (Dynamic - system GMM) for GLBD} \label{tab:Dynamic - system GMM_ger} \medskip
\renewcommand{\arraystretch}{1}
\setlength{\tabcolsep}{14pt}
\begin{tabular}{l|c| c}
\toprule
\textbf{Independent Variables}&\textbf{GLBD} &\textbf{GSynLBD}\\
\midrule
AR(1) Coefficient&   0.1883\sym{***}&   0.6140\sym{***}\\
          & (0.0021)         & (0.0027)         \\
[1em]
Ln Pay    &   0.6599\sym{***}&   0.3553\sym{***}\\
          & (0.0014)         & (0.0020)         \\
[1em]
Age 3-4   &  -0.0292\sym{***}&  -0.0934\sym{***}\\
          & (0.0011)         & (0.0013)         \\
[1em]
Age 5-7   &  -0.0512\sym{***}&  -0.1266\sym{***}\\
          & (0.0011)         & (0.0014)         \\
[1em]
Age 8-12  &  -0.0791\sym{***}&  -0.1545\sym{***}\\
          & (0.0011)         & (0.0015)         \\
[1em]
Age 13 or more&  -0.1400\sym{***}&  -0.2012\sym{***}\\
          & (0.0011)         & (0.0015)         \\
\hline
\(N\)     &   848871         &   966084         \\
m2        &  19.4900         &  -8.8300         \\
Sargan test&  4.5e+03         &  2.8e+03         \\
df of Sargan Test& 526.0000         & 526.0000         \\
P value of Sargan test&   0.0000         &   0.0000         \\
    \bottomrule
  \end{tabular} 
\begin{tablenotes}
\small
\item Note: An observation is an entity-year. In this table, $m2$ is the Arellano-Bond test for zero autocorrelation in first-differenced errors for order two. Standard errors are in parentheses. ***, **, and * indicate statistically significant coefficients at 1\%, 5\%, and 10\% percent levels, respectively.
 \end{tablenotes}
\end{table}

\begin{table}[H]
  \centering
\begin{threeparttable}
 \caption{Regression coefficients (Dynamic - system GMM with MA(1)) for LEAP} \label{tab:Dynamic - system GMM with MA(1)_can} \medskip
\renewcommand{\arraystretch}{1}
\begin{tabular}{l|c c| c c}
\toprule
\textbf{Independent Variables}&\multicolumn{2}{c|}{\textbf{LEAP}} &  \multicolumn{2}{c}{\textbf{CanSynLBD}}\\
\midrule
&\multicolumn{1}{c}{Private}&\multicolumn{1}{c}{Manufacturing}&\multicolumn{1}{c}{Private}&\multicolumn{1}{c}{Manufacturing}\\
\hline
AR(1) Coefficient&   0.2005\sym{***}&   0.2821\sym{***}&   0.4850\sym{***}&   0.5737\sym{***}\\
          & (0.0007)         & (0.0040)         & (0.0012)         & (0.0059)         \\
[1em]
Ln Pay    &   0.8044\sym{***}&   0.7135\sym{***}&   0.4760\sym{***}&   0.4056\sym{***}\\
          & (0.0005)         & (0.0034)         & (0.0009)         & (0.0046)         \\
[1em]
Age 3-4   &  -0.1245\sym{***}&  -0.2033\sym{***}&  -0.1716\sym{***}&  -0.3158\sym{***}\\
          & (0.0005)         & (0.0032)         & (0.0006)         & (0.0037)         \\
[1em]
Age 5-7   &  -0.1328\sym{***}&  -0.2264\sym{***}&  -0.1733\sym{***}&  -0.3389\sym{***}\\
          & (0.0005)         & (0.0035)         & (0.0006)         & (0.0043)         \\
[1em]
Age 8-12  &  -0.1383\sym{***}&  -0.2454\sym{***}&  -0.1731\sym{***}&  -0.3560\sym{***}\\
          & (0.0006)         & (0.0039)         & (0.0007)         & (0.0051)         \\
[1em]
Age 13 or more&  -0.1441\sym{***}&  -0.2586\sym{***}&  -0.1774\sym{***}&  -0.3717\sym{***}\\
          & (0.0006)         & (0.0042)         & (0.0008)         & (0.0058)         \\
\hline
\(N\)     & 15708195         &  1015293         & 13573225         &   959764         \\
m2        &   8.2000         &   7.0600         & -40.0300         &  -6.6400         \\
Sargan test&  2.8e+04         &  2.3e+03         &  1.7e+04         &  1.3e+03         \\
df of Sargan Test& 251.0000         & 251.0000         & 251.0000         & 251.0000         \\
P value of Sargan test&   0.0000         &   0.0000         &   0.0000         &   0.0000         \\
    \bottomrule
  \end{tabular} 
\begin{tablenotes}
\small
\item Note: An observation is a firm and a year. In this table, $m2$ is the Arellano-Bond test for zero autocorrelation in first-differenced errors for order two.  Standard errors are in parentheses. ***, **, and * indicate statistically significant coefficients at 1\%, 5\%, and 10\% percent levels, respectively.
 \end{tablenotes}
 \end{threeparttable}
\end{table}

\begin{table}[H]
  \centering
\caption{Regression coefficients (Dynamic - system GMM with MA(1)) for GLBD} \label{tab:Dynamic - system GMM with MA(1)_ger} \medskip
\renewcommand{\arraystretch}{1}
\setlength{\tabcolsep}{14pt}
\begin{tabular}{l|c| c}
\toprule
\textbf{Independent Variables}&\textbf{GLBD} &\textbf{GSynLBD}\\
\midrule
AR(1) Coefficient&   0.3701\sym{***}&   0.5268\sym{***}\\
          & (0.0060)         & (0.0048)         \\
[1em]
Ln Pay    &   0.5349\sym{***}&   0.4202\sym{***}\\
          & (0.0041)         & (0.0036)         \\
[1em]
Age 3-4   &  -0.0594\sym{***}&  -0.0831\sym{***}\\
          & (0.0015)         & (0.0013)         \\
[1em]
Age 5-7   &  -0.0922\sym{***}&  -0.1105\sym{***}\\
          & (0.0018)         & (0.0015)         \\
[1em]
Age 8-12  &  -0.1252\sym{***}&  -0.1351\sym{***}\\
          & (0.0019)         & (0.0016)         \\
[1em]
Age 13 or more&  -0.1850\sym{***}&  -0.1802\sym{***}\\
          & (0.0019)         & (0.0017)         \\
\hline
\(N\)     &   848871         &   966084         \\
m2        &  19.0300         & -11.6900         \\
Sargan test&  3.1e+03         &  2.5e+03         \\
df of Sargan Test& 494.0000         & 494.0000         \\
P value of Sargan test&   0.0000         &   0.0000         \\
    \bottomrule
  \end{tabular} 
\begin{tablenotes}
\small
\item Note: An observation is a firm and a year. In this table, $m2$ is the Arellano-Bond test for zero autocorrelation in first-differenced errors for order two. Standard errors are in parentheses. ***, **, and * indicate statistically significant coefficients at 1\%, 5\%, and 10\% percent levels, respectively.
 \end{tablenotes}
\end{table}

\section{Canada: Synthesized Observations}
\label{sec:synth_obs}

\begin{table}[H]
  \centering
\begin{threeparttable}
  \caption{Synthesized observations}  \label{tab:Synthesized_observations} \medskip
  \renewcommand{\arraystretch}{1}
  \begin{tabular}{l  c c }
    \toprule
    \textbf{Category}&\textbf{\# of Observations (millions)}&\textbf{Percentage}\\
    \midrule
Synthesized&22.01&93.35\\
Not synthesized&1.57&6.65\\
Total&23.58&100.00\\
    \bottomrule
  \end{tabular} 
\begin{tablenotes}
\small
\item Note: Industries that are not synthesized are NAICS 4481,    4482,     4483,     4511,     4513,     4841,     4842, 5241, and 5242. We drop observations from  synthesized industries when there are  less than ten observations in a given year. We do not synthesize the public sector (NAICS 61, 62, and 91).
 \end{tablenotes}
 \end{threeparttable}
\end{table}
 
\section{Confidentiality assessment}
\label{sec:conf:appendix}

\begin{table}[H]
\centering\footnotesize
\caption{Observed entity births given synthetic births for LEAP.} \label{tab:Can:ProbabilityPrivate} \medskip
\renewcommand{\arraystretch}{1}
\begin{tabular}{c c| c c c}
\toprule
\multicolumn{2}{c|}{\textbf{First (Birth) Year}} &  \multicolumn{3}{c}{\textbf{\% of Births over NAICS}}\\
\textbf{Synthetic}&\textbf{Actual}&\textbf{Minimum}&\textbf{Mean}&\textbf{Maximum}\\
\midrule
1991&1991&0.00&27.69&83.02\\
1992&1992&0.00&3.37&11.11\\
1993&1993&0.00&3.79&33.33\\
1994&1994&0.00&3.73&33.33\\
1995&1995&0.00&3.86&20.00\\
1996&1996&0.00&4.25&33.33\\
1997&1997&0.00&4.10&16.94\\
1998&1998&0.00&4.41&25.00\\
1999&1999&0.00&4.23&33.33\\
2000&2000&0.00&3.41&25.00\\
2001&2001&0.00&2.73&22.22\\
2002&2002&0.00&2.65&25.00\\
2003&2003&0.00&2.22&10.00\\
2004&2004&0.00&2.60&17.86\\
2005&2005&0.00&2.71&20.00\\
2006&2006&0.00&2.83&50.00\\
2007&2007&0.00&2.90&33.33\\
2008&2008&0.00&2.38&20.00\\
2009&2009&0.00&2.47&50.00\\
2010&2010&0.00&2.12&33.33\\
2011&2011&0.00&2.65&50.00\\
2012&2012&0.00&2.41&20.00\\
2013&2013&0.00&2.48&25.00\\
2014&2014&0.00&2.23&20.00\\
2015&2015&0.00&2.15&33.33\\
 \bottomrule
\end{tabular} 
\\
\justify
\end{table}

\begin{table}[H]
\centering\footnotesize
\caption{Observed entity births given synthetic births (GLBD)} \label{tab:GLBD:Probability} \medskip
\renewcommand{\arraystretch}{1}
\begin{tabular}{c c| c c c}
\toprule
\multicolumn{2}{c|}{\textbf{Birth Year}} &  \multicolumn{3}{c}{\textbf{\% of Births over NAICS}}\\
\textbf{Synthetic}&\textbf{Actual}&\textbf{Minimum}&\textbf{Mean}&\textbf{Maximum}\\
\midrule
1976&1976&18.34&19.77&21.20\\
1977&1977&1.35&1.55&1.75\\
1978&1978&0.97&1.50&2.02\\
1979&1979&1.99&2.05&2.11\\
1980&1980&1.15&1.61&2.07\\
1981&1981&0.76&1.28&1.80\\
1982&1982&1.29&1.39&1.48\\
1983&1983&1.54&1.57&1.61\\
1984&1984&0.99&1.03&1.07\\
1985&1985&0.83&1.56&2.28\\
1986&1986&1.36&1.79&2.21\\
1987&1987&1.99&2.00&2.02\\
1988&1988&1.18&1.49&1.81\\
1989&1989&1.65&1.84&2.03\\
1990&1990&2.44&2.79&3.14\\
1991&1991&7.59&9.17&10.75\\
1992&1992&5.19&8.81&12.42\\
1993&1993&3.20&3.40&3.60\\
1994&1994&3.50&3.93&4.35\\
1995&1995&2.86&3.26&3.65\\
1996&1996&1.89&2.62&3.35\\
1997&1997&3.46&3.96&4.45\\
1998&1998&3.58&3.68&3.78\\
1999&1999&5.56&5.78&6.00\\
2000&2000&3.19&3.64&4.10\\
2001&2001&3.26&3.59&3.93\\
2002&2002&2.04&3.00&3.97\\
2003&2003&2.13&3.17&4.20\\
2004&2004&2.57&3.24&3.91\\
2005&2005&1.66&2.54&3.41\\
2006&2006&2.15&3.06&3.97\\
2007&2007&2.17&2.90&3.62\\
2008&2008&2.37&2.42&2.47\\
 \bottomrule
\end{tabular} 
\\
\justify
\end{table}